\documentclass[aps,prd,12pt]{article}
\usepackage{graphicx}
\usepackage{graphics,epsfig}
\usepackage{amssymb,amsmath,bm}
\usepackage{hyperref} 

\usepackage[T1]{fontenc}
\usepackage[utf8]{inputenc}
\usepackage{authblk}

\usepackage{slashed}
\usepackage{pifont}

\renewcommand{\thefootnote}{\fnsymbol{footnote}}
\newcommand{\as}{\alpha_s}

\newcommand{\ul}[1]{\underline{#1}}

\newcommand{\ord}[1]{\mathcal{O}\left(#1\right)}
\newcommand{\Tr}{\mathrm{Tr}}
\def\eq#1{{Eq.~(\ref{#1})}}
\def\fig#1{{Fig.~\ref{#1}}}

\def\ru1{\rule[-0.4truecm]{0mm}{1truecm}}
\title{\vspace*{1cm} \Large \bf Sivers Function in the Quasi-Classical
  Approximation}

\author{Yuri V. Kovchegov}
\author{Matthew D. Sievert}

\affil{\small Department of Physics, The Ohio State University, Columbus,
  OH 43210, USA}

\date{\today}

\begin{document}

\maketitle

\thispagestyle{empty}

\begin{abstract}
  We calculate the Sivers function in semi-inclusive deep inelastic
  scattering (SIDIS) and in the Drell-Yan process (DY) by employing
  the quasi-classical Glauber--Mueller/ McLerran--Venugopalan
  approximation. Modeling the hadron as a large ``nucleus'' with
  non-zero orbital angular momentum (OAM), we find that its Sivers
  function receives two dominant contributions: one contribution is
  due to the OAM, while another one is due to the local Sivers
  function density in the nucleus. While the latter mechanism, being
  due to the ``lensing'' interactions, dominates at large transverse
  momentum of the produced hadron in SIDIS or of the di-lepton pair in
  DY, the former (OAM) mechanism is leading in saturation power
  counting and dominates when the above transverse momenta become of
  the order of the saturation scale. We show that the OAM channel
  allows for a particularly simple and intuitive interpretation of the
  celebrated sign flip between the Sivers functions in SIDIS and DY.
\end{abstract}

~\\
\centerline{PACS numbers: 12.38.Bx, 13.88.+e, 12.38.Cy, 24.85.+p}

\thispagestyle{empty}

\setcounter{page}{0}

\newpage

\setcounter{footnote}{0}
\renewcommand{\thefootnote}{\arabic{footnote}}


\section{Introduction}

Single transverse spin asymmetries (STSAs) generated in semi-inclusive
deep inelastic scattering (SIDIS) and in hadronic collisions are one
of the hot topics of research in quantum chromodynamics (QCD),
promising unparalleled insight in the physics of chiral symmetry
breaking and quark confinement. In the factorization framework
involving transverse momentum-dependent distribution functions (TMDs)
\cite{Collins:1989gx,Collins:1981uk} the origin of STSAs is chiefly
ascribed to either the quark TMDs (Sivers effect \cite{Sivers:1989cc,
  Sivers:1990fh}), to multiple partonic rescattering
\cite{Efremov:1981sh,Efremov:1984ip,Qiu:1991pp,Ji:1992eu,Qiu:1998ia,Brodsky:2002cx,Collins:2002kn,Koike:2011mb,Kanazawa:2000hz,Kanazawa:2000kp},
or to the quark fragmentation functions (Collins effect
\cite{Collins:1992kk}).

While both the quark TMD and the fragmentation function are
non-perturbative, and, according to the conventional wisdom, cannot
be calculated form first principles, it is desirable to understand the
detailed physical mechanism leading to generation of STSAs in QCD. To
that end a significant progress has been achieved by Brodsky, Hwang
and Schmidt (BHS) in \cite{Brodsky:2002cx} (see also
\cite{Kane:1978nd,Efremov:1981sh,Efremov:1984ip,Qiu:1991pp,Qiu:1998ia,Collins:2002kn,Kovchegov:2012ga}),
where, in a quark--di-quark proton model calculation, it has been
shown that the STSA in SIDIS can be generated through an interference
of the final-state parton rescattering diagram with the Born-level
amplitude. In essence, it was shown in
\cite{Brodsky:2002cx,Collins:2002kn} that multiple partonic
rescatterings are key to generating the asymmetry. The multiple
rescatterings are often referred to as the ``lensing'' interaction,
since, in SIDIS, the associated color-Lorentz force tries to attract
the knocked-out quarks back into the hadron
\cite{Burkardt:2008ps,Burkardt:2010sy}, thus ``focusing'' them. The
effects of such multiple rescatterings can be absorbed into the Sivers
distribution function of a polarized hadron in SIDIS
\cite{Collins:2002kn,Belitsky:2002sm}.

A consequence of this understanding of the origin of STSA in SIDIS, is
that the Sivers function (and, hence, the asymmetry itself) has to
change sign between SIDIS and the Drell-Yan process (DY). At the level
of the operator matrix element this conclusion has been reached in
\cite{Collins:2002kn}, while an illustration of this result in the BHS
model was completed only recently \cite{Brodsky:2013oya} (see also
\cite{Brodsky:2002rv} for the outline of the calculation). It is our
understanding that in the ``lensing'' interpretation of STSAs this
sign change corresponds to the color-Lorentz force changing sign from
attractive to repulsive between a knocked-out quark in SIDIS and the
incoming anti-quark in DY.

The goal of the present work is to extend our understanding of the
physical mechanism behind the STSA beyond the quark--di-quark model of
the proton used in
\cite{Brodsky:2002cx,Brodsky:2002rv,Brodsky:2013oya} (see
\cite{Gamberg:2010xi,Gamberg:2009uk} for other efforts in a similar
direction). In particular, multiple partonic rescatterings in high
energy scattering can be particularly simply accounted for in the
framework of the quasi-classical approximation to QCD employed in the
Glauber--Mueller (GM) \cite{Mueller:1989st} and, equivalently,
McLerran--Venugopalan (MV)
\cite{McLerran:1993ka,McLerran:1994vd,McLerran:1993ni} models. In
these approaches the hadron is modeled by a large nucleus, with a
large number $A$ of nucleons in it. The large number of nucleons leads
to high density of small-$x$ gluons in the nuclear wave function,
which, in turn, generates a hard scale $Q_s \gg \Lambda_{QCD}$ known
as the parton saturation scale, justifying the use of perturbative QCD
calculations. (For reviews of the saturation/Color Glass Condensate
(CGC) physics see
\cite{Iancu:2003xm,Weigert:2005us,Jalilian-Marian:2005jf,Gelis:2010nm,KovchegovLevin}.)
The fact that the quasi-classical approximation generates a hard scale
justifying the approach indicates that it is not simply a ``model'' of
QCD, but, in fact, it represents a limiting behavior of strong
interactions at high energy. Multiple rescatterings can be resummed in
the GM/MV model as an expansion in powers of the parameter $\as^2 \,
A^{1/3}$ \cite{Kovchegov:1997pc}: the presence of a resummation
parameter allows for a controlled approximation to the problem at
hand. In the past there was a number of efforts to include spin
effects in the saturation/CGC framework
\cite{Boer:2006rj,Boer:2008ze,Boer:2002ij,Dominguez:2011br,Metz:2011wb,Kovchegov:2012ga,Kang:2011ni,Kang:2012vm}.

To alleviate the worry about whether a large nucleus can adequately
represent a proton (or any other hadron) in SIDIS and DY experiments,
let us point out that in unpolarized scattering the proton may have a
significant number of non-perturbatively--generated large-$x$ ($x >
0.01$) partons, which are modeled by ``nucleons'' in this
large-nucleus approximation. The large-$x$ partons/''nucleons'', in
turn, give rise to small-$x$ gluons. The resulting expressions for the
deep inelastic scattering (DIS) structure functions have been quite
successful in describing HERA low-$x$ data
\cite{Golec-Biernat:1998js,Golec-Biernat:1999qd,Kowalski:2003hm,Tribedy:2010ab},
also indicating relevance of the large-''nucleus'' approximation to
the proton wave function at small-$x$.

In what follows we would have to slightly modify the original MV model
of the nucleus by giving the ``nucleus'' both a non-zero spin and a
non-zero orbital angular momentum (OAM). Here this would mean that
free nucleons in an approximately spherical bag, as considered
originally in \cite{McLerran:1993ka,McLerran:1994vd,McLerran:1993ni},
would now be polarized and would be orbiting the nuclear spin axis. In
a realistic polarized nucleus the nucleons tend to form pairs with
zero net OAM, such that the net spin of the nucleus is carried by the
few unpaired nucleons and does not get very large (does not grow
directly with $A$). Since it is not clear whether such effect (at the
level of quarks and gluons) takes place in the proton we are trying to
model, we will not make any particular assumptions about the
polarizations and OAMs of the nucleons in our ``nucleus''.

The main physical mechanism for generating STSA in the quasi-classical
framework is as follows. Imagine a large spinning nucleus. The nucleus
is so large that it is almost completely opaque to a colored
probe. This strong nuclear shadowing is due to multiple rescatterings
in the nucleus generating a short mean free path for the quark,
anti-quark, or a gluon.

\begin{figure}[h]
\centering
\includegraphics[height=4cm]{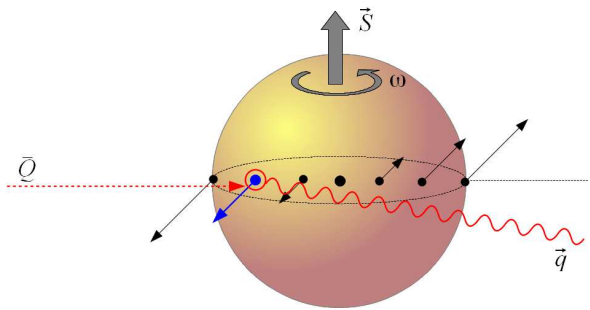}
\caption{The physical mechanism of STSA in DY as envisioned in the
  text.}
\label{DY_fig}
\end{figure}

Let us first consider the Drell-Yan process on such a rotating nucleus
with shadowing, as shown in \fig{DY_fig} in the nuclear rest frame
with the rotation axis of the nucleus perpendicular to the collision
axis. The incoming anti-quark (generated in the wave function of the
other hadron) scatters on the ``front'' surface of the polarized
nucleus due to the strong shadowing. Since the anti-quark interacts
with the nucleons which, at the ``front'' of the nucleus
preferentially rotate with the nucleus out of the plane of the page in
\fig{DY_fig}, the produced time-like virtual photons are produced
preferentially out of the page, generating left-of-polarized-beam
single spin asymmetry.\footnote{This mechanism is similar in spirit to
  the original way of thinking by D. Sivers about the single
  transverse spin asymmetry. (D. Sivers to M. Sievert, private
  communications.) A heuristic classical picture of a polarized hadron
  or nucleus was pioneered in \cite{Chou:1976jf}.}

\begin{figure}[h]
\centering
\includegraphics[height=4cm]{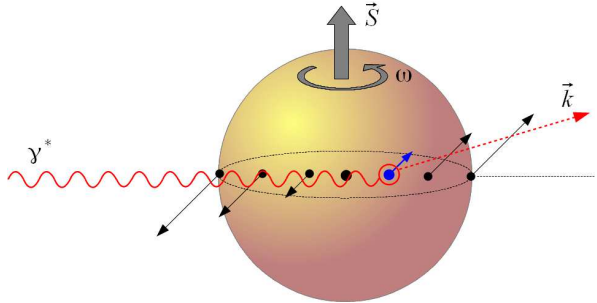}
\caption{The physical mechanism of STSA in SIDIS as envisioned in the
  text.}
\label{SIDIS_fig}
\end{figure}

The same mechanism can be applied to generate STSA in SIDIS, as
illustrated in \fig{SIDIS_fig}, also in the rest frame of the
nucleus. Now the incoming virtual photon interacts with the
transversely polarized nucleus, producing a quark. For the quark to
escape out of the nucleus and be produced the interaction has to take
place at the ``back'' of the nucleus, to minimize the path the quark
needs to travel through the nucleus, maximizing its chances to
escape. The nucleons in the ``back'' of the nucleus rotate
preferentially into the page of \fig{SIDIS_fig}: scattering of a
virtual photon on such nucleons results in the right-of-beam single
spin asymmetry for the outgoing quarks (quarks produced preferentially
with transverse momentum pointing into the page).

Spin asymmetries in DY and SIDIS shown in Figs.~\ref{DY_fig} and
\ref{SIDIS_fig} are generated through a combination of OAM effects and
nuclear shadowing. The two asymmetries are opposite-sign (left- and
right-of-beam), and, assuming that scattering in the two processes
happens equal distance from the nuclear edge, are likely to be equal
in magnitude, in agreement with the prediction of
\cite{Collins:2002kn,Brodsky:2002rv}. 

As we will see below in the actual calculations, the STSAs in
Figs.~\ref{DY_fig} and \ref{SIDIS_fig} do require multiple
rescatterings, but they are needed solely to generate nuclear
shadowing. Thus the physical mechanism of Figs.~\ref{DY_fig} and
\ref{SIDIS_fig} is quite different from the ``lensing'' interaction
\cite{Brodsky:2002cx,Brodsky:2002rv}, in which the knocked-out quark
in SIDIS ``feels'' the net color charge of the remainder of the
proton, and is attracted back by this charge
\cite{Burkardt:2008ps,Burkardt:2010sy}.\footnote{Applying this logic
  to DY one would expect that to obtain an STSA sign reversal compared
  to SIDIS one needs the anti-quark in DY to ``feel'' an equal
  repulsive force from the rest of proton (that is, from the proton
  without the quark which is about to annihilate the antiquark):
  however, it is unclear to us how the incoming anti-quark can
  ``feel'' the force of only a part of the intact proton (excluding
  the quark) while interacting with the whole proton coherently.} In
the presence of shadowing, it would be much harder for the quark in
SIDIS to ``see'' the whole remainder of the polarized proton (nucleus)
coherently: thus one expects the ``lensing'' effect to weaken with
increasing shadowing (if we could increase shadowing without modifying
the degree of the polarization of the nucleus). This is qualitatively
different from the mechanism in Figs.~\ref{DY_fig} and
\ref{SIDIS_fig}, in which the asymmetry actually increases with
shadowing. Clearly, the more opaque the nucleus is, the more likely
the interactions to happen at its ``front'' in DY and at its ``back''
in SIDIS, making the asymmetry larger.

In the paper below we will outline the calculations leading to the
physical picture presented in Figs.~\ref{DY_fig} and
\ref{SIDIS_fig}. After some generalities in Sec.~\ref{sec:general} we
proceed in Sec.~\ref{sec:SIDIS} with the quasi-classical analysis of
STSA in SIDIS. As mentioned above, to model the OAM of a polarized
nucleus we have to assume that the nucleus is rotating. This implies a
generalization of the original MV and GM models, in which nucleons are
static, to include rotational motion of the nucleons. Hence the
nucleons need to have both the well-defined positions and momenta:
this is only possible in the classical limit. The classical MV model
limit is achieved in Sec.~\ref{sec:SIDIS} using the Wigner functions
approach, which allows to specify both the positions and momenta of
the nucleons in the polarized nucleus. 

We then proceed to the calculation of STSA in SIDIS, identifying two
mechanisms for STSA generation: one is due to the coupling of the
produced quarks transverse momentum to the OAM of the nucleus, while
another one is due to the STSA generated in the scattering of the
virtual photon on an individual nucleon along the lines of the BHS
mechanism \cite{Brodsky:2002cx} (Sivers function density). The former
mechanism is leading in the saturation framework, being dominant in
the saturation power counting (for non-zero OAM): it is order-one for
$\as^2 \, A^{1/3} \sim 1$. The latter mechanism is order-$\as$ for
$\as^2 \, A^{1/3} \sim 1$, and is thus subleading. 

At large values of the produced quark transverse momentum $k_T$ the
OAM mechanism gives the contribution to the Sivers function of the
order $A \, \as \, m_N \, p_T \, Q_s^2 / k_T^6$ with $p_T$ the typical
transverse momentum of the valence quarks in the polarized nucleus due
to orbital motion and $m_N$ the nucleon mass (with $m_N/3$ roughly the
constituent quark mass), while the Sivers function density gives a
contribution proportional to $A \, \as^2 \, m_N^2/k_T^4$. Assuming
that $p_T \approx m_N$, we see that the Sivers function density
mechanism dominates for $k_T > Q_s/\sqrt{\as}$; conversely, the OAM
mechanism is dominant for $k_T < Q_s/\sqrt{\as}$, the domain including
everything inside of the saturation region and a phase-space sector
outside of that region.

A similar quasi-classical STSA calculation is carried out for the
Drell-Yan process in Sec.~\ref{sec:DY}, where we also explicitly show
the mechanism for the sign reversal of the Sivers function outlined in
this Introduction. We conclude in Sec.~\ref{sec:Discussion} by
summarizing our results and outlining possible improvements of our
results left for the future work.


\section{Definitions: Single Spin Asymmetries, Sivers Function}
\label{sec:general}

The single transverse spin asymmetry is defined as 
\begin{align}
 \label{eq-Defn STSA}
 A_N ({\ul k}) \equiv \; \frac{\frac{d \sigma^\uparrow}{d^2 k \, dy} -
   \frac{d \sigma^\downarrow}{d^2 k \, dy}} {\frac{d
     \sigma^\uparrow}{d^2 k \, dy} + \frac{d \sigma^\downarrow}{d^2 k
     \, dy}} \; = \; \frac{\frac{d \sigma^\uparrow}{d^2 k \, dy} (\ul
   k) - \frac{d \sigma^\uparrow}{d^2 k \, dy} (- \ul k)} {\frac{d
     \sigma^\uparrow}{d^2 k \, dy} (\ul k) + \frac{d
     \sigma^\uparrow}{d^2 k \, dy} (- \ul k)}
\end{align}
for producing a hadron with transverse momentum $\ul k$ in SIDIS on a
transversely polarized target and in polarized proton--proton
collisions or a di-lepton pair with transverse momentum $\ul k$ in DY
process on a polarized proton. The asymmetry $A_N$ singles out a part
of the production cross section proportional to $({\vec S} \times
{\vec p} \, ) \cdot {\vec k}$, where $\vec p$ is the 3-momentum of the
polarized hadron pointing along the collision axis.

Throughout this paper we will use light-cone coordinates $p^\pm \equiv
p^0 \pm p^3$ with the corresponding metric $p \cdot q = \tfrac{1}{2}
p^+ q^- + \tfrac{1}{2} p^- q^+ - \ul{p} \cdot \ul{q}$.  Accordingly,
we denote four-vectors as $p^\mu = (p^+ , p^- , \ul{p})$, with the
transverse momentum $\ul{p} \equiv (p^1 , p^2)$ and $p_T = p_\perp =
|\ul{p}|$.

As we have outline above, a possible physical explanation of the
asymmetry is the Sivers effect \cite{Sivers:1989cc,
  Sivers:1990fh}. The aim of this work is to calculate the Sivers
function in the quasi-classical approximation. To define the Sivers
function first consider a quark-quark correlation function in a
polarized hadron or nucleus defined by
\cite{Boer:2011xd,Boer:2002ju}
\begin{align}
  \label{eq:q_corr}
  \Phi_{ij} (x, {\ul k}; P, S) \equiv \int \frac{d x^- \, d^2
    x_\perp}{2 (2 \, \pi)^3} \, e^{i \, \left(\frac{1}{2} \, x \, P^+ \,
      x^- - {\ul x} \cdot {\ul k} \right)} \, \langle P, S | {\bar
    \psi}_j (0) \, {\cal U} \, \psi_i (x^+=0, x^-, {\ul x}) | P, S
  \rangle ,
\end{align}
where $\psi_i$ is the quark field with Dirac index $i = 1, \ldots, 4$,
while the quark is taken with transverse momentum ${\ul k}$ and the
longitudinal momentum fraction $x$. The proton (or polarized nucleus)
spin four-vector is $S^\mu$, while ${\cal U}$ is the gauge link
necessary to make the object on the right of \eq{eq:q_corr}
gauge-invariant.

Below, when considering SIDIS and DY, we will work in the light-cone
gauge of the projectile. Choosing the polarized proton (nucleus) to
move along the light-cone $x^+$-direction, such that $P^+$ is large,
we will work in the $A^- =0$ gauge. In the quasi-classical
approximation the $A^- =0$ gluon field of a large ultrarelativistic
nucleus moving along the $x^+$-direction has zero transverse
component, ${\ul A} =0$, such that the only non-zero component is
$A^+$. Defining the Wilson line 
\begin{align} 
  \label{Wlines1}
  V_{\ul x} [b^- \, , \, a^-] \equiv \mathcal{P} \, \exp \left[
    \frac{i g}{2} \int\limits_{a^-}^{b^-} dx^- A^+ (x^+ = 0, x^-,
    \ul{x}) \right]
\end{align}
we write for the case of SIDIS \cite{Collins:2002kn,Belitsky:2002sm}
\begin{align}
  \label{eq:U_SIDIS}
  {\cal U}^{SIDIS} = V_{\ul 0}^\dagger [+\infty \, , \, 0] \, V_{\ul
    x} [+\infty \, , \, x^-],
\end{align}
while for DY we have 
\begin{align}
  \label{eq:U_DY}
  {\cal U}^{DY} = V_{\ul 0} [0 \, , \, -\infty ] \, V_{\ul x}^\dagger
  [x^- \, , \, -\infty ].
\end{align}
In both cases we neglected the transverse gauge link at $x^- = \pm
\infty$ since ${\ul A} =0$ in the gauge we chose. As will become
apparent below, the direction of the Wilson lines in the $\cal U$'s is
given by the direction of motion of the outgoing quark in SIDIS and
the incoming anti-quark in DY. This results in different definitions
of the correlator $\Phi_{ij}$ for the two processes, which is usually
referred to as the controlled process-dependence of the TMDs
\cite{Collins:2002kn}.

The correlation function $\Phi_{ij}$ is decomposed as
\cite{Boer:1997nt,Boer:2002ju}
\begin{align}
  \label{eq:Phi_dec}
  & \Phi_{ij} (x, {\ul k}; P, S) = \frac{M}{2 \, P^+} \, \bigg[ f_1
  (x, k_T) \, \frac{\slashed{P}}{M} + \frac{1}{M^2} \, f_{1 \,
    T}^\perp (x, k_T) \, \epsilon_{\mu\nu\rho\sigma} \, \gamma^\mu \,
  P^\nu \, k_\perp^\rho \, S_\perp^\sigma - \frac{1}{M} \, g_{1s} (x,
  \ul{k}) \, \slashed{P} \, \gamma^5 \notag \\ & - \frac{1}{M} \,
  h_{1T} (x, k_T) \, i \, \sigma_{\mu\nu} \, \gamma^5 \, S_\perp^\mu
  \, P^\nu - \frac{1}{M^2} \, h_{1s}^\perp (x, \ul{k}) \, i \,
  \sigma_{\mu\nu} \, \gamma^5 \, k_\perp^\mu \, P^\nu + h_{1}^\perp
  (x, k_T) \, \sigma_{\mu\nu} \, \frac{k_\perp^\mu \, P^\nu}{M^2}
  \bigg]_{ij},
\end{align}
where $M$ is the mass of the polarized proton or nucleus. 

In the following we will be using the Sivers function $f_{1 \,
  T}^\perp (x, {\ul k})$ and the unpolarized quark TMD $f_1 (x, {\ul
  k})$. These functions can be obtained from the correlator
$\Phi_{ij}$ using the following expressions
\begin{subequations}\label{TMDs}
  \begin{align}
    \label{eq:quark_TMD}
    \Phi_{ij} (\gamma^+)_{ji} \bigg|_{\text{spin independent}} = 2 \,
    f_{1} (x,  k_T);
  \end{align}
\begin{align}
  \label{eq:Sivers_ext}
  \Phi_{ij} (\gamma^+)_{ji} \bigg|_{\text{spin dependent}} =
  \frac{2}{M} \, \epsilon^{ij} \, S_\perp^i \, k_\perp^j \, f_{1 \,
    T}^\perp (x, k_T).
\end{align}
\end{subequations}


\section{Semi-Inclusive Deep Inelastic Scattering}
\label{sec:SIDIS}

We first consider the process of quark production in semi-inclusive
deep inelastic lepton scattering on a transversely polarized heavy
nucleus: $\ell + A^\uparrow \rightarrow \ell' + q + X$.  The leptonic
tensor can be factorized out in the usual way, so we represent the
process as the scattering of a virtual photon: $\gamma^* + A^\uparrow
\rightarrow q + X$.  This photon carries a large spacelike virtuality
$q_\mu q^\mu = -Q^2$ and knocks out a quark from one of the nucleons,
which may then rescatter on the nuclear remnants. The nucleus is taken
in the classical GM/MV approximation, which we augment by requiring
that the nucleons are polarized and the nucleus rotates around the
transverse spin axis, which leads to a non-zero OAM.

\begin{figure}[h]
\centering
\includegraphics[height=4cm]{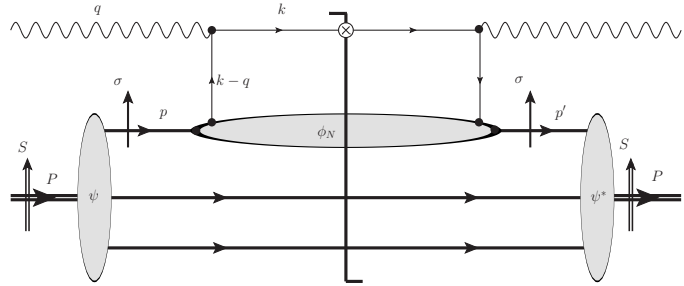}
\caption{The lowest-order SIDIS process in the usual $\alpha_s$
  power-counting. A quark is ejected from a nucleon in the nucleus by
  the high-virtuality photon, which escapes without
  rescattering. Different solid horizontal lines represent valence
  quarks from different nucleons in the nuclear wave function, with
  the latter denoted by the vertical shaded oval.}
\label{fig:DIS1}
\end{figure}

Consider first the lowest-order process shown in Fig.~\ref{fig:DIS1},
in which a quark is ejected without rescattering.\footnote{In
  small-$x$ physics quark production is dominated by a higher-order in
  $\as$ process, where the virtual photon splits into a $q\bar q$ pair
  before hitting the target: since in this work $x = \ord{1}$, the
  dipole process is not dominant, constituting an order-$\as$
  correction to the channel shown in \fig{fig:DIS1}.} We work in a
frame (such as the photon-nucleus center-of-mass frame) in which the
virtual photon moves along the $x^-$-axis with a large momentum $q^-$
and the nucleus moves along the $x^+$-axis with a large momentum
$P^+$.  In this frame, the kinematics are
\begin{align} \label{DIS1}
 \begin{aligned}
   P^\mu &= \left( P^+ , \frac{M_A^2}{P^+}, \ul{0} \right) \\
   q^\mu &= \left( -\frac{Q^2}{q^-} , q^- , \ul{0} \right) \\
   p^\mu &= \left( \alpha P^+ , \frac{p_T^2 + m_N^2}{\alpha P^+} , \ul{p} \right) \\
   k^\mu &= \left( \frac{k_T^2}{k^-} , k^- , \ul{k} \right),
 \end{aligned}
\end{align}
where $M_A$ is the mass of the nucleus and the on-mass-shell valence
quark with momentum $p^\mu$ is a part of the light-cone wave function
of the nucleus. In what follows we will model nucleons as made out of
single valence quarks: in the end of the calculation, to go back to
the nucleons one simply would need to replace distribution functions
in a valence quark by the distribution functions in the nucleons.

Let us denote the photon-nucleus center-of-mass energy squared by $s_A
\equiv(P+q)^2$ and the photon-nucleon (valence quark) center-of-mass
energy squared by $\hat{s} \equiv (p+q)^2$.  We consider the kinematic
limit $s_A \gg \hat{s} , Q^2 \gg p_T^2 , k_T^2 , M_A^2$ and work to
leading order in the small kinematic quantities
$\tfrac{\bot^2}{\hat{s}} , \tfrac{\bot^2}{Q^2}$, which we denote
collectively as $\mathcal{O}(\tfrac{\bot^2}{Q^2})$.  Since we are 
operating in the limit in which $Q^2 \gg \bot^2 \gg \Lambda^2$, the
formalism of TMD factorization applies, justifying the use of the 
correlator \eqref{eq:q_corr} and decomposition \eqref{eq:Phi_dec}.
Additionally, to
a good accuracy one can assume that a typical scale for the momentum
fraction $\alpha$ is $\ord{1/A}$, where $A$ is the mass number of the
nucleus (in fact, $\alpha \approx 1/A$ for the single-valence quark
"nucleons" at hand).  In this limit,
\begin{align} \label{DIS2}
 \begin{aligned}
   p^+ q^- &= \hat{s} + Q^2 \\
   q^+ &= - \left( \frac{Q^2}{\hat{s}+Q^2} \right) p^+ = - x \, p^+ =
   - \alpha \, x \, P^+
 \end{aligned}
\end{align}
where $x \equiv Q^2 / (2 p \cdot q)$ is the Bjorken scaling variable
per nucleon.  The corresponding scaling variable for the entire
nucleus is $x_A \equiv Q^2 / (2 P \cdot q) = \alpha \, x \approx x/A$.
The kinematic limit at hand, $\hat{s} \sim Q^2 \gg p_T^2 , k_T^2 ,
M_A^2$ corresponds to $x \sim \ord{1}$.  The on-shell condition for
the outgoing gluon is
\begin{align} 
  \label{DIS3} k^- = \frac{k_T^2}{k^+} = q^- + \frac{p_T^2 +
    m_N^2}{\alpha P^+} - \frac{(\ul{p}-{\ul k})_T^2}{\alpha P^+ -
    \alpha \, x \, P^+ - k^+} \approx q^-
\end{align}
which fixes the struck quark to be ejected along the $x^-$-direction,
so that its light-cone plus momentum
\begin{align} 
  \label{DIS4} k^+ = \frac{k_T^2}{q^-} = \left(\frac{k_T^2}{\hat{s} +
      Q^2}\right) p^+ = \left(\frac{k_T^2}{Q^2} \right) \alpha \, x \,
  P^+
\end{align}
is small since $\sqrt{\hat{s}} \sim Q \sim p^+ \gg k_T$.  This also
fixes the momentum fraction of the active quark just before
interaction with the photon to be $x_F \equiv (k^+ - q^+)/p^+ \approx
- q^+/p^+ = x$ in the usual way. (Note that $q^+ = - Q^2/q^- < 0$.)

In our frame, the $x^-$-extent of the Lorentz-contracted nucleus is
$L^- \sim \tfrac{M_A}{P^+} R$, where $R$ is the radius of the nucleus
in its rest frame.  The incoming virtual photon and outgoing quark
interact with the nucleus based on their corresponding coherence
lengths: $\ell_\gamma^- \sim 1/|q^+|$ and $\ell_k^- \sim 1/k^+$,
respectively.  Comparing these to the size of the nucleus,
\begin{align} \label{DIS5}
 \begin{aligned}
   \frac{\ell_\gamma^-}{L^-} &\sim \frac{1}{x} \frac{1}{\alpha M_A R}
   \sim \ord{A^{-1/3}} \ll 1,
   \\
   \frac{\ell_k^-}{L^-} &\sim \frac{1}{x}
   \left(\frac{Q^2}{k_T^2}\right) \frac{1}{\alpha M_A R} \sim
   \ord{\frac{Q^2 + \hat{s}}{\bot^2} A^{-1/3}} \gg 1,
 \end{aligned}
\end{align}
we see that the photon's coherence length is short, but the coherence
length of the ejected quark is parametrically large for $\hat{s}, Q^2
\gg \perp^2 \, A^{1/3}$.  Thus, for our calculation in which $x \sim
\ord{1}$, the virtual photon interacts incoherently (locally) on a
single nucleon, but the ejected quark interacts coherently with all of
the remaining nucleons it encounters before escaping the nucleus.

This limit thus combines the local ``knockout'' picture of the deep
inelastic scattering process with the coherent rescattering that
usually characterizes the small-$x$ limit.  In the formal limit of a
large nucleus in which $\alpha_s \ll 1$ and $A \gg 1$ such that
$\alpha_s^2 A^{1/3} \sim \ord{1}$, these coherent interactions with
subsequent nucleons must be re-summed according to this
saturation-based power counting.


\subsection{Quark Production in SIDIS}

In general it is rather straightforward to write an answer for the
quasi-classical quark production in SIDIS. As we mentioned in the
Introduction, here the problem is a little more subtle than usual
since we are interested in also including transverse and longitudinal
motion of the nucleons in the nucleus in order to model its OAM. Thus
our quasi-classical description of the nucleus has to provide us both
with the positions and momenta of the nucleons. This can be done using
Wigner distributions.

Let us illustrate the method with a simple single-rescattering process
from \fig{fig:DIS1}. Just like in the parton model, the time scale of
inter-nucleon interactions is Lorentz-dilated in the infinite momentum
frame of the nucleus that we are working in. We can, therefore, write
the scattering amplitude for the process in \fig{fig:DIS1} as a
product of the light-cone wave function $\psi$ of the valence quarks
in the nucleus (defined according to light-front perturbation theory
rules \cite{Lepage:1980fj,Brodsky:1983gc} in the boost-invariant
convention of \cite{KovchegovLevin}) with the quark--virtual photon
scattering amplitude $M_K$:
\begin{align}
  \label{eq:wf_LO}
  M_{tot} = \psi (p) \, M_K (p, q, k). 
\end{align}
Here $\psi (p) = \psi (p^+/P^+, {\ul p})$ is the boost-invariant
light-cone wave function of a valence quark (in one of the nucleons)
in the nucleus, while $M_K$ is the scattering amplitude for the
``knock-out'' process $\gamma^* + q \to q + X$. The sum over valence
quark spin and color is implied in \eqref{eq:wf_LO}. In calculating
the quark production process we need to square this amplitude,
integrate it over the momentum of the final state gluon and sum over
all nucleons in the nucleus. Since momenta $k$ and $q$ are fixed, this
amounts to integrating over $p$. One gets
\begin{align}
  \label{eq:ampl2_LO}
  \int \frac{d p^+ \, d^2 p}{2 (p^++ q^+) \, (2 \pi)^3} \, |M_{tot}|^2
  = A \, \int \frac{d p^+ \, d^2 p}{2 (p^+ + q^+) \, (2 \pi)^3} \,
  |\psi (p)|^2 \, |M_K (p, q, k)|^2.
\end{align}

First let us introduce a Fourier transform of the valence quark wave
function,
\begin{align}
  \label{eq:Fourier}
  \psi (b) \equiv \psi (b^-, {\ul b}) = \int \frac{d p^+ d^2 p}{2 \, \sqrt{p^+} \,
    (2 \pi)^3} \, e^{- i \, p \cdot b} \, \psi (p),
\end{align}
with $p \cdot b = \tfrac{1}{2} \, p^+ \, b^- - {\ul p} \cdot {\ul b}$.
Next we define the Wigner distribution for the valence quarks (which
is closely related to the Wigner distribution of the nucleons in the
quasi-classical MV model employed here) with the help of the Fourier
transform \eqref{eq:Fourier}:
\begin{align} 
  \label{DIS8} 
  W(p,b) \equiv W (p^+, {\ul p}; b^- , {\ul b}) \equiv \int d^2 \delta
  b \, d \delta b^- \, e^{i \, p \cdot \delta b} \, \psi(b +
  \tfrac{1}{2} \delta b) \, \psi^*(b - \tfrac{1}{2} \delta b).
\end{align}
Note that the wave function is normalized such that
\begin{align}
\int \frac{d p^+ d^2 p}{2 \, p^+ \,
    (2 \pi)^3} \, |\psi (p)|^2 =1
\end{align}
giving
\begin{align} \label{rr2}
  \int  \frac{d p^+ \, d^2 p \, d b^- \, d^2 b}{2 (2\pi)^3} \, W(p,b) = 1.
\end{align}
Since
\begin{align}
\int d^2 b \, d b^- \, W (p,b) = |\psi (p)|^2 /p^+
\end{align}
we can recast \eq{eq:ampl2_LO} as
\begin{align}
  \label{eq:ampl2_LO_W}
  \int \frac{d p^+ \, d^2 p}{2 (p^++ q^+) \, (2 \pi)^3} \, |M_{tot}|^2
  = A \, \int \frac{d p^+ \, d^2 p \, d b^- \, d^2 b}{2 \, (2 \pi)^3}
  \, W(p,b) \, \frac{p^+}{p^+ + q^+} \, |M_K (p, q, k)|^2.
\end{align}

Finally, in the following, as usual in the saturation framework, it
would be convenient to calculate the scattering amplitude in (partial)
transverse coordinate space. Writing
\begin{align}
\label{M_Ftr}
M_K (p, q, k) = \int d^2 x \, e^{-i \, {\ul k} \cdot ({\ul x} - {\ul
    b})} \, M_K (p,q, {\ul x} - {\ul b} )
\end{align}
(with $k^-$ and $k^+$ fixed by Eqs.~\eqref{DIS3} and \eqref{DIS4}) we
rewrite \eq{eq:ampl2_LO_W} as
\begin{align}
  \label{eq:ampl2_LO_Wxy}
  \int \frac{d p^+ \, d^2 p}{2 (p^++ q^+) \, (2 \pi)^3} \, |M_{tot}|^2
  = A \, \int \frac{d p^+ \, d^2 p \, d b^- \, d^2 b}{2 \, (2 \pi)^3}
  \, W(p,b) \, \frac{p^+}{p^+ + q^+} \notag \\ \times \, \int d^2x \,
  d^2 y \, e^{-i \, {\ul k} \cdot ({\ul x} - {\ul y})} \, M_K (p, q,
  {\ul x} - {\ul b} ) \, M_K^* (p, q, {\ul y} - {\ul b} ).
\end{align}

Note that the Fourier transform \eqref{M_Ftr} appears to imply that
${\ul b}$ is the transverse position of the outgoing gluon in
\fig{fig:DIS1}, whereas in the Wigner distribution ${\ul b}$ is the
position of the valence quark $p$. As we will shortly see such
interpretation is not inconsistent: in the classical limit of a large
nucleus the Wigner distribution is a slowly varying function of $\ul
b$, with changes in $W$ becoming significant over the variations of
$\ul b$ over distances of the order of nucleon size 1 fm or
larger. The valence quark and outgoing gluon in \fig{fig:DIS1} are
perturbatively close to each other (being the part of the same Feynman
diagram), and hence the difference in their positions is outside the
precision of $W (p,b)$ and can be taken to be the same in the Wigner
distribution.

In Appendix~\ref{A} we show that the formula \eqref{eq:ampl2_LO_Wxy}
holds not only at the lowest order, but when multiple rescatterings are
included as well, such that in the kinematics outlined above
\begin{align}
  \label{eq:ampl2_LO_Wxy_net}
  \int \frac{d p^+ \, d^2 p}{2 (p^++ q^+) \, (2 \pi)^3} \, |A_{tot}|^2
  = A \, \int \frac{d p^+ \, d^2 p \, d b^- \, d^2 b}{2 \, (2 \pi)^3}
  \, W(p,b) \, \frac{p^+}{p^+ + q^+} \notag \\ \times \, \int d^2x \,
  d^2 y \, e^{-i \, {\ul k} \cdot ({\ul x} - {\ul y})} \, A (p, q,
  {\ul x} - {\ul b} ) \, A^* (p, q, {\ul y} - {\ul b} ),
\end{align}
where we define the energy-independent (at the lowest non-trivial
order) $2 \to 2$ scattering amplitudes by (see also
Eqs.~\eqref{eq:e-ind_ampl} and \eqref{eq:A_Eindep})
\cite{KovchegovLevin}
\begin{align}
  \label{eq:AM}
  A (p, q, k) = \frac{M (p, q, k)}{2 \, p^+ \, q^-}
\end{align}
and $A(p, q, k)$ in \eq{eq:ampl2_LO_Wxy_net} denotes the sum over
rescatterings of the virtual photon on any number of nucleons in the
nucleus.\footnote{Strictly-speaking we need to include in
  \eq{eq:ampl2_LO_Wxy_net} Wigner function convolutions with the all
  the interacting nucleons in the nucleus: however, since in our
  kinematics only the first ``knockout'' process depends on the
  transverse momentum $p_\perp$ of the nucleon, we only keep one
  convolution with the Wigner function explicitly.} (Note that for a
``nucleus'' made out of a single nucleon $p^+ = P^+$, which allows one
to reduce \eq{eq:ampl2_LO_Wxy} to \eq{eq:ampl2_LO_Wxy_net} by
neglecting the ``spectator'' nucleons.) We therefore conclude that the
quark production cross section for the $\gamma^* + A \to q + X$
process can be written as
\begin{align}\label{xsect_W}
  \frac{d \sigma^{\gamma^* + A \to q + X}}{d^2 k \, dy} = A \, \int
  \frac{d p^+ \, d^2 p \, d b^- \, d^2 b}{2 \, (2 \pi)^3} \, W(p,b) \,
  \frac{d \hat{\sigma}^{\gamma^* + NN\ldots N \to q + X}}{d^2 k \,
    dy},
\end{align}
where the cross section for producing a quark in $\gamma^*$ scattering
on the nucleons is
\begin{align}\label{xsectNNN}
  \frac{d \hat{\sigma}^{\gamma^* + NN\ldots N \to q + X}}{d^2 k \, dy}
  = {\cal N} \, \int d^2x \, d^2 y \, e^{-i \, {\ul k} \cdot ({\ul x}
    - {\ul y})} \, A_K (p, q, {\ul x} - {\ul b} ) \, A_K^* (p, q, {\ul
    y} - {\ul b} ) \, D_{{\ul x} \, {\ul y}} [+\infty, b^-]
\end{align}
with the semi-infinite fundamental dipole scattering amplitude given
by (cf. \eq{eq:U_SIDIS})
\begin{align}
 \label{dipole_def}
  D_{{\ul x} \, {\ul y}} [+\infty, b^-] = \left\langle \frac{1}{N_c}
    \, \mbox{Tr} \left[ V_{\ul x} [+\infty, b^-] \, V^\dagger_{\ul y}
      [+\infty, b^-] \right] \right\rangle
\end{align}
and with some ${\hat s}$ and $Q^2$-dependent prefactor $\cal N$. Here
$y = \ln 1/x$ is the rapidity of the produced quark and a factor of
$A$ in \eq{xsect_W} accounts for the fact that the first scattering
can take place on any of the $A$ nucleons. We fixed the normalization
of \eq{xsect_W} by requiring it to be valid for a nucleus made out of
a single nucleon, which would be described by a trivial Wigner
distribution fixing the momentum and position of the nucleon by simple
delta-functions. (Alternatively one could require the formula to be
valid for the case of cross section $\hat \sigma$ independent of $p$
and $b$.)

As already mentioned before, with the accuracy of the large-$A$
classical approximation, the argument $\ul b$ in the Wigner
distribution can be replaced by any other transverse coordinate
involved in the scattering process. Hence one can replace $\ul b$ in
$W (b,p)$ from \eq{xsect_W} by either $\ul x$ or $\ul y$ from
\eq{xsectNNN}, or by any linear combination of those
variables. Replacing $\ul b$ in $W (b,p)$ from \eq{xsect_W} by $({\ul
  x} + {\ul y})/2$ and employing \eq{xsectNNN} we write
\begin{align}\label{xsect_W2}
  \frac{d \sigma^{\gamma^* + A \to q + X}}{d^2 k \, dy} = & A \, \int
  \frac{d p^+ \, d^2 p \, d b^- }{2 (2 \pi)^3} \, \int d^2x \, d^2 y \
  W \bigg( p , b^-, \frac{{\ul x} + {\ul y}}{2} \bigg) \notag \\ &
  \times \, e^{- i \, {\ul k} \cdot ({\ul x} - {\ul y})} \, |A_K|^2
  (p, q, {\ul x} - {\ul y} ) \, D_{{\ul x} \, {\ul y}} [+\infty, b^-],
\end{align}
where
\begin{align}
\label{MK2}
|A_K|^2 (p, q, {\ul x} - {\ul y} ) & \equiv \, {\cal N} \, \int d^2 b
\, A_K (p, q, {\ul x} - {\ul b} ) \, A_K^* (p, q, {\ul y} - {\ul b} )
\notag \\ & = \int \frac{d^2 k'}{(2 \pi)^2} \, e^{i \, {\ul k}' \cdot
  ({\ul x} - {\ul y})} \, \frac{d \hat{\sigma}^{\gamma^* + N \to q +
    X}}{d^2 k' \, dy} (p,q).
\end{align}
Substituting \eq{MK2} into \eq{xsect_W2} yields
\begin{align}\label{xsect_W3}
  \frac{d \sigma^{\gamma^* + A \to q + X}}{d^2 k \, dy} = & A \, \int
  \frac{d p^+ \, d^2 p \, d b^- }{2 (2 \pi)^3} \, \int d^2x \, d^2 y \
  W \bigg( p , b^-, \frac{{\ul x} + {\ul y}}{2} \bigg) \notag \\ &
  \times \int \frac{d^2 k'}{(2 \pi)^2} \, e^{- i \,({\ul k} - {\ul
      k}') \cdot ({\ul x} - {\ul y})} \, \frac{d
    \hat{\sigma}^{\gamma^* + N \to q + X}}{d^2 k' \, dy} (p,q) \,
  D_{{\ul x} \, {\ul y}} [+\infty, b^-].
\end{align}
\eq{xsect_W3} is our starting point for exploring the STSA in SIDIS:
it gives the quark production cross section in the quasi-classical
approximation.

\begin{figure}[t]
\centering
\includegraphics[width= .5 \textwidth]{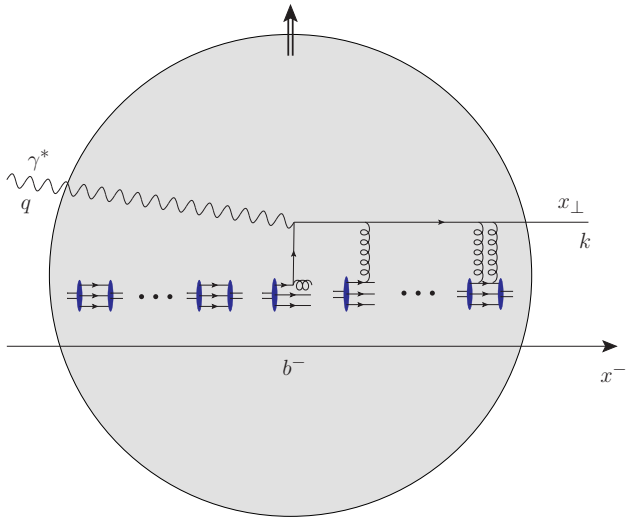}
\caption{Space-time structure of quark production in the
  quasi-classical SIDIS process in the rest frame of the nucleus,
  overlaid with one of the corresponding Feynman diagrams. The shaded
  circle is the transversely polarized nucleus, with the vertical
  double arrow denoting the spin direction.}
\label{SIDIS_qprod}
\end{figure}

The expression \eqref{xsect_W3} is illustrated in \fig{SIDIS_qprod}:
the first interaction between the incident virtual photon and a
nucleon in the transversely polarized nucleus happens at the
longitudinal coordinate $b^-$. A quark is knocked out, which proceeds
to interact with the rest of the nucleons in the nucleus. This latter
interaction is recoilless and is encoded in a Wilson line.

The Wigner distribution in \eq{xsect_W3} allows to take the
quasi-classical GM/MV limit of a large nucleus in a controlled
way. For a large "classical" nucleus we usually can replace $W(p,b)$
by the following classical expression for it (neglecting longitudinal
orbital motion of the nucleons)
\begin{align}
\label{Wcl}
W_{cl} (p,b) = \frac{4 \, \pi}{A} \, \rho ({\ul b}, b^-) \, \delta
\left( p^+ - \frac{P^+}{A} \right) \, w ( {\ul p} , b),
\end{align}
where $\rho ({\ul b}, b^-)$ is the nucleon number density normalized
such that
\begin{align}
\int d^2 b \, d b^- \rho ({\ul b}, b^-) = A. 
\end{align}
The function $w ( {\ul p} , b)$ in \eq{Wcl} is responsible for the
transverse momentum distribution of the nucleons and, to satisfy
\eq{rr2}, is normalized such that
\begin{align}
\int \frac{d^2 p}{(2 \pi)^2} \, w ( {\ul p} , b) =1. 
\end{align}
As originally formulated
\cite{McLerran:1993ka,McLerran:1994vd,McLerran:1993ni}, the MV model
contained no dependence on the spin or transverse momentum of the
valence quarks.  This result is recovered by using $w_{MV} = (2 \pi)^2
\, \delta^2 ({\ul p})$.

Substituting the classical Wigner distribution \eqref{Wcl} into
\eq{xsect_W3} yields
\begin{align}\label{xsect_W4}
  \frac{d \sigma^{\gamma^* + A \to q + X}}{d^2 k \, dy} = & \int
  \frac{d^2 p \, d b^- }{(2 \pi)^2} \, d^2x \, d^2 y \ \rho \left
    (\frac{{\ul x} + {\ul y}}{2}, b^- \right) \, w \bigg( {\ul p} ,
  \frac{{\ul x} + {\ul y}}{2}, b^- \bigg) \notag \\ & \times \, \int
  \frac{d^2 k'}{(2 \pi)^2} \, e^{- i \,({\ul k} - {\ul k}') \cdot
    ({\ul x} - {\ul y})} \, \frac{d \hat{\sigma}^{\gamma^* + N \to q +
      X}}{d^2 k' \, dy} (p,q) \, D_{{\ul x} \, {\ul y}} [+\infty,
  b^-],
\end{align}
which is a simplified version of \eq{xsect_W3}.


\subsection{Quasi-Classical Sivers Function in SIDIS}

Imagine a large nucleus with the total spin $\vec J$ such that
\begin{align}
  \label{eq:spin_dec}
  {\vec J} = {\vec L} + {\vec S},
\end{align}
where ${\vec L}$ is the OAM of all the nucleons in the nucleus and
${\vec S}$ is the net spin of all the nucleons. In the quasi-classical
approximation at hand the OAM is generated by rotation of the nucleons
around a preferred axis. The nucleus is transversely polarized to the
beam: we assume that both ${\vec L}$ and ${\vec S}$ point along the
(positive or negative) $\hat x$-axis.

The result \eqref{xsect_W3} for the quark production cross section in
SIDIS can be utilized to write down an expression for SIDIS Sivers
function of the large nucleus with the help of \eq{eq:Sivers_ext}. We
first note that the quark production cross section in SIDIS is
proportional to the correlator \eqref{eq:q_corr} with the
future-pointing Wilson line given by \eq{eq:U_SIDIS}
(cf. Eqs.~\eqref{xsectNNN} and \eqref{dipole_def}).  The gauge link in
\eqref{dipole_def} begins and ends at the same $b^-$, while the more
general gauge link in \eqref{eq:U_SIDIS} has different endpoints at
$0$ and $x^-$.  The reason is that the nuclear wave function is
composed of color-neutral ``nucleons'' localized in $b^-$; hence there
is only a contribution to the correlator when the gauge link both
begins and ends at the same $b^-$.  The Dirac $\gamma^+$-matrix of
\eq{eq:Sivers_ext} is also present in the quark production cross
section since the Dirac structure of the large-$k^-$ outgoing quark
line is given by $\gamma^+ \, k^-$. To obtain the Sivers function one
only needs to eliminate the gamma--matrices stemming from the
quark--photon vertices in the amplitude and in the complex conjugate
amplitude; this can be done by simply contracting the Lorentz indices
of these gamma--matrices \cite{Brodsky:2013oya}. While such
contraction is not allowed in a calculation of the SIDIS cross section
due to non-trivial structure of the lepton tensor, it is a legitimate
method of extracting the Sivers function \cite{Brodsky:2013oya}, since
$\gamma_\mu \, \gamma^+ \, \gamma^\mu = -2 \, \gamma^+$. We thus see
that and equation like \eqref{xsect_W3} would still hold for Tr$[\Phi
\, \gamma^+]$ instead of SIDIS cross section, since to obtain the
former one simply needs to repeat all the steps of the cross section
derivation that led to \eq{xsect_W3} without inserting the photon
polarizations (implicit in \eqref{xsect_W3}), and adding a contraction
over Lorentz indices of the gamma--matrices from the quark--photon
vertices in the end.

\begin{figure}[ht]
\centering
\includegraphics[width= .75 \textwidth]{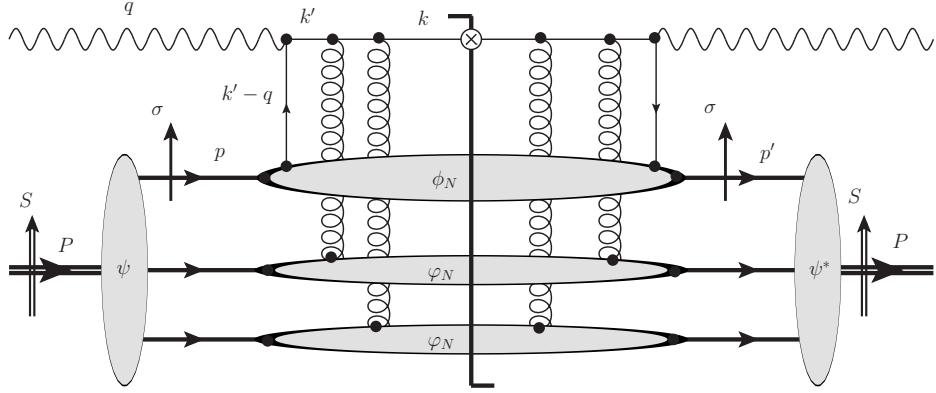}
\caption{
  \label{fig:TMD1} Decomposition of the nuclear quark distribution
  $\Phi_A$ probed by the SIDIS virtual photon into mean-field wave
  functions $\psi, \psi^*$ of nucleons and the quark and gluon
  distributions $\phi_N$ and $\varphi_N$ of the nucleons.  }
\end{figure}

By analogy with \eq{xsect_W3} we can express the quark correlation
function $\Phi_A$ of the nucleus in terms of the quasi-classical
distribution $W_N(p,b)$ of nucleons, the quark correlators $\phi_N$ of
individual nucleons, and the semi-infinite Wilson line trace $D_{{\ul
    x} {\ul y}}[+\infty, b^-]$:
\begin{align} 
 \label{TMD1}
 \Tr[\Phi_A({\bar x},\ul{k}; P, J) \, \gamma^+] = A \int \frac{d p^+
   \, d^2 p \, d b^-}{2(2\pi)^3} &\, d^2x \, d^2 y \, \sum_\sigma
 W_N^\sigma \left( p,b^-, \frac{{\ul x} + {\ul y}}{2} \right) \int
 \frac{d^2 k'}{(2\pi)^2} \, e^{ - i \, (\ul{k} - \ul{k'})
   \cdot(\ul{x}-\ul{y})} \notag \\ &\times \Tr[\phi_N(x ,\ul{k}' - x
 \, {\ul p}; p , \sigma) \, \gamma^+] \, D_{{\ul x} {\ul y}}[+\infty,
 b^-].
\end{align}
\eq{TMD1} is illustrated in \fig{fig:TMD1}. In \eq{TMD1} we explicitly
show the sum over the polarizations $\sigma = \pm 1/2$ of the nucleons 
along the $x$-axis. Note that $x = -
q^+/p+$ and it varies with $p^+$ inside the integral; at the same time
the ``averaged'' value of Bjorken-$x$ per nucleon is ${\bar x} = - A
\, q^+/P^+$.  The quark correlator of the nucleus $\Phi_A$ is defined
by \eq{eq:q_corr},
\begin{align}
  \label{TMD2}
  \Phi_{ij}^A ({\bar x}, {\ul k}; P, J) \equiv \int \frac{d x^- \, d^2
    x_\perp}{2(2 \, \pi)^3} \, e^{i \, \left(\frac{1}{2} \, {\bar x} \,
      P^+ \, x^- - {\ul x} \cdot {\ul k} \right)} \, \langle A; P, J|
  {\bar \psi}_j (0) \, {\cal U}^{SIDIS} \, \psi_i (x^+=0, x^-, {\ul
    x}) | A; P, J \rangle ,
\end{align}
along with the corresponding nucleonic correlator
is
\begin{align} 
  \label{TMD3} 
  \phi_{ij}^N (x, {\ul k}; p, \sigma) \equiv \int \frac{d x^- \, d^2
    x_\perp}{2(2 \, \pi)^3} \, e^{i \, \left(\frac{1}{2} \, x \, p^+ \,
      x^- - {\ul x} \cdot {\ul k} \right)} \, \langle N; p, \sigma |
  {\bar \psi}_j (0) \, {\cal U}^{SIDIS} \, \psi_i (x^+=0, x^-, {\ul
    x}) | N; p, \sigma \rangle.
\end{align}
These definitions are done in a frame in which the parent particle's
transverse momentum is zero. The $\ul{k}' - x \, {\ul p}$ in the
argument of $\phi_N$ in \eq{TMD1} is obtained by making a transverse
boost from the frame in which the nucleon has transverse momentum
$\ul{p}$ into a frame in which $\ul{p}=\ul{0}$ (and the definition
\eqref{TMD3} applies).  Note that our lab frame corresponds to the
photon-nucleus center-of-mass frame in which $\ul{q} = \ul{P} =
\ul{0}$. The polarization-dependent Wigner functions are normalized as
(cf. \eq{rr2})
\begin{align} 
\begin{aligned}
  \label{Wnorm} \int \frac{d p^+ \, d^2 p \, d b^- \, d^2 b}{2
    (2\pi)^3} \, A \, W^{+1/2} (p,b) & = \# \ \mbox{spin-up nucleons} \, ; \\
  \int \frac{d p^+ \, d^2 p \, d b^- \, d^2 b}{2
    (2\pi)^3} \, A \, W^{-1/2} (p,b) & = \# \ \mbox{spin-down nucleons} \, .
\end{aligned}
\end{align}

As in \cite{Boer:2002ju}, the correlation functions in \eq{TMD1} can
be parametrized in terms of the TMD distribution functions, of which
the most relevant to the problem at hand are the unpolarized
distribution $f_1$ and the Sivers function $f_{1T}^\perp$. Using
Eqs.~\eqref{TMDs} we write
\begin{subequations}\label{TMD4}
\begin{align} 
  \Tr[\Phi_A ({\bar x}, \ul{k}; P, J) \, \gamma^+] &= 2 \, f_1^A
  ({\bar x}, k_T) + \, \frac{2}{M_A} \, {\hat z} \cdot (\ul{J}
  \times \ul{k})\, f_{1T}^{\perp A} ({\bar x}, k_T) \label{TMD4A} \\
  \Tr[\phi_N (x , \ul{k}'- x \, \ul{p}; p, \sigma) \, \gamma^+] &= 2
  \, f_1^N (x, |\ul{k}'- x \, \ul{p}|) + \notag \\ &+ \, \frac{2}{m_N}
  \, {\hat z} \cdot \left( \ul{\sigma} \times (\ul{k}'- x \, \ul{p})
  \right) \ f_{1T}^{\perp N} (x, |\ul{k}'- x \,
  \ul{p}|), \label{TMD4B}
\end{align}
\end{subequations}
where we introduced the unpolarized quark TMDs ($f_1^A$ and $f_1^N$)
and Sivers functions ($f_{1T}^{\perp A}$ and $f_{1T}^{\perp N}$) for
the nucleus and nucleons respectively, along with the masses $M_A$ and
$m_N$ of the nucleus and nucleons.

We may extract the Sivers function of the nucleus $f_{1T}^{\perp A}$
by antisymmetrizing \eqref{TMD4A} with respect to either the nuclear
spin or the momentum $\ul{k}$ of the produced quark:\footnote{In doing
  so we assume that the Sivers function is an even function of
  $\ul{k}$, which is indeed the case due to its $T$-symmetry
  properties.}
\begin{align} 
  \label{TMD5} {\hat z} \cdot (\ul{J}\times\ul{k}) \, f_{1T}^{\perp A}
  ({\bar x} , k_T) = \frac{1}{4} M_A \, \Tr[\Phi_A ({\bar x},\ul{k}; P
  , J) \, \gamma^+] - (\ul{k} \rightarrow -\ul{k}).
\end{align}
Using \eq{TMD1} in \eq{TMD5} we write
\begin{align} \label{TMD6}
 \begin{aligned}
   {\hat z} \cdot & (\ul{J}\times\ul{k}) \, f_{1T}^{\perp A} ({\bar x}
   , k_T) = \frac{1}{4} M_A \, A \int \frac{d p^+ \, d^2 p \, d
     b^-}{2(2\pi)^3} \, d^2x \, d^2 y \, \sum_\sigma W_N^\sigma \left(
     p,b^-,
     \frac{{\ul x} + {\ul y}}{2} \right) \\
   & \times \int \frac{d^2 k'}{(2\pi)^2} \, e^{ - i \, (\ul{k} -
     \ul{k'}) \cdot(\ul{x}-\ul{y})} \, \Tr[\phi_N(x ,\ul{k}' - x \,
   {\ul p}; p , \sigma) \, \gamma^+] \, D_{{\ul x} {\ul y}}[+\infty,
   b^-] - (\ul{k} \rightarrow - \ul{k}).
 \end{aligned}
\end{align}
We can decompose the quark correlator in a nucleon $\phi_N$ into the
nucleon's unpolarized quark distribution $f_1^N$ and Sivers function
$f_{1T}^N$ using \eqref{TMD4B}. Substituting this into \eq{TMD6}
yields
\begin{align} \label{TMD7}
 \begin{aligned}
   {\hat z} \cdot & (\ul{J}\times\ul{k}) \, f_{1T}^{\perp A} ({\bar x}
   , k_T) = \frac{1}{4} M_A \, A \int \frac{d p^+ \, d^2 p \, d
     b^-}{2(2\pi)^3} \, d^2x \, d^2 y \, \sum_\sigma W_N^\sigma \left(
     p,b^-,
     \frac{{\ul x} + {\ul y}}{2} \right)  \int \frac{d^2 k'}{(2\pi)^2} \\
   & \times \, e^{ - i \, (\ul{k} - \ul{k'}) \cdot(\ul{x}-\ul{y})} \,
   \left[ 2 \, f_1^N (x, |\ul{k}'- x \, \ul{p}|) + \frac{2}{m_N} \,
     {\hat z} \cdot \left( \ul{\sigma} \times (\ul{k}'- x \, \ul{p})
     \right) \ f_{1T}^{\perp N} (x, |\ul{k}'- x \, \ul{p}|) \right] \\
   & \times \, D_{{\ul x} {\ul y}}[+\infty, b^-] - (\ul{k} \rightarrow
   - \ul{k}).
 \end{aligned}
\end{align}

We can understand the sources of the $T$-odd nuclear Sivers function
$f_{1T}^{\perp A}$ by explicitly (anti)symmetrizing the various terms
on the right of \eq{TMD7}. To start with, perform the nucleon spin sum
$\sum_\sigma$ in a basis parallel or antiparallel to the nuclear spin
$\ul{S}$. This can be done using the definitions
\begin{align} \label{TMD8}
 \begin{aligned}
   \sum_\sigma W_N^\sigma(p,b) & \equiv W_{unp}(p,b) \\
   \sum_\sigma W_N^\sigma(p,b) \, \ul{\sigma}& \equiv \frac{1}{A} \,
   W_{trans}(p,b) \, \ul{S},
 \end{aligned}
\end{align}
where we will refer to $W_{unp}$ as the distribution of unpolarized
nucleons and to $W_{trans}$ as the nucleon transversity
distribution. Note that 
\begin{align} 
  \label{Wnorm2} \int \frac{d p^+ \, d^2 p \, d b^- \, d^2 b}{2
    (2\pi)^3} \, W_{unp} (p,b) = 1, \ \ \ \int \frac{d p^+ \, d^2 p \,
    d b^- \, d^2 b}{2 (2\pi)^3} \, W_{trans} (p,b) = 1,
\end{align}
as follows from the definition \eqref{TMD8} and from \eqref{Wnorm}.

\eq{TMD7} becomes
\begin{align} 
  \label{TMD9} {\hat z} \cdot & (\ul{J}\times\ul{k}) \, f_{1T}^{\perp
    A} ({\bar x} , k_T) = \frac{M_A}{2} \int \frac{d p^+ \, d^2 p \, d
    b^-}{2(2\pi)^3} \, d^2x \, d^2 y \, \frac{d^2 k'}{(2\pi)^2} \, e^{
    - i \, (\ul{k} - \ul{k'}) \cdot(\ul{x}-\ul{y})} \, \bigg[ A \,
  W_{unp} \left( p,b^-, \frac{{\ul x} +
      {\ul y}}{2} \right) \notag \\
  & \times \, f_1^N (x, |\ul{k}'- x \, \ul{p}|) + W_{trans} \left(
    p,b^-, \frac{{\ul x} + {\ul y}}{2} \right) \frac{1}{m_N} \, {\hat
    z} \cdot \left( \ul{S} \times (\ul{k}'- x \, \ul{p})
  \right) \, f_{1T}^{\perp N} (x, |\ul{k}'- x \, \ul{p}|) \bigg] \notag \\
  & \times \, D_{{\ul x} {\ul y}}[+\infty, b^-] - (\ul{k} \rightarrow
  - \ul{k}).
\end{align}
Now, in the terms with $(\ul{k} \rightarrow -\ul{k})$ being
subtracted, we also redefine the dummy integration variables $\ul{x}
\leftrightarrow \ul{y}$, $\ul{k'} \rightarrow - \ul{k'}$, and $\ul{p}
\rightarrow - \ul{p}$.  This leaves the Fourier factors and the
distribution functions $f_{1}^{N}$, $f_{1 T}^{\perp \, N}$ unchanged,
giving
\begin{align} 
  \label{TMD10} {\hat z} \cdot & (\ul{J}\times\ul{k}) \, f_{1T}^{\perp
    A} ({\bar x} , k_T) = \frac{M_A}{2} \int \frac{d p^+ \, d^2 p \, d
    b^-}{2(2\pi)^3} \, d^2x \, d^2 y \, \frac{d^2 k'}{(2\pi)^2} \, e^{
    - i \, (\ul{k} - \ul{k'}) \cdot(\ul{x}-\ul{y})} \, \bigg\{
  f_1^N (x, |\ul{k}'- x \, \ul{p}|) \notag \\
  & \times \, A \, \left[ W_{unp} \left( p^+, \ul{p} ,b^-, \frac{{\ul
          x} + {\ul y}}{2} \right) \, D_{{\ul x} {\ul y}}[+\infty,
    b^-] - W_{unp} \left( p^+ , -\ul{p},b^-, \frac{{\ul x} + {\ul
          y}}{2} \right) \, D_{{\ul y} {\ul x}}[+\infty, b^-] \right]
  \notag \\ & + \frac{1}{m_N} \, {\hat z} \cdot \left( \ul{S} \times
    (\ul{k}'- x \,
    \ul{p}) \right) \, f_{1T}^{\perp N} (x, |\ul{k}'- x \, \ul{p}|) \\
  & \times \, \left[ W_{trans} \left( p^+, \ul{p} ,b^-, \frac{{\ul x}
        + {\ul y}}{2} \right) \, D_{{\ul x} {\ul y}}[+\infty, b^-] +
    W_{trans} \left( p^+, - \ul{p} ,b^-, \frac{{\ul x} + {\ul y}}{2}
    \right) \, D_{{\ul y} {\ul x}}[+\infty, b^-] \right]
  \bigg\}. \notag
\end{align}
At this point it is convenient to explicitly (anti)symmetrize the
distribution functions with respect to $\ul{p} \leftrightarrow
-\ul{p}$ and the Wilson lines with respect to $\ul{x} \leftrightarrow
\ul{y}$. Define
\begin{align} \label{Wlines3}
 \begin{matrix}
  S_{{\ul x} {\ul y}} \equiv \tfrac{1}{2}(D_{{\ul x} {\ul y}} + D_{{\ul y} {\ul x}}) \\
  i \, O_{{\ul x} {\ul y}} \equiv \tfrac{1}{2}(D_{{\ul x} {\ul y}} - D_{{\ul y} {\ul x}})
 \end{matrix}
 \hspace{2cm}
 D_{{\ul x} {\ul y}} = S_{{\ul x} {\ul y}} + i \, O_{{\ul x} {\ul y}}
\end{align}
and
\begin{align} \label{DIS21}
 \begin{aligned}
   W^{\left( \substack{symm \\ OAM} \right)}(p,b) \equiv \tfrac{1}{2}
   \left[ W(p,b) \pm (\ul{p} \rightarrow - \ul{p}) \right],
 \end{aligned}
\end{align}
where we have used the ``OAM'' label to indicate that the preferred
direction of transverse momentum in the antisymmetric case reflects
the presence of net orbital angular momentum.  We can decompose $W$
into symmetric and OAM parts for both the unpolarized distribution
$W_{unp}$ and the transversity distribution $W_{trans}$.

Using the (anti)symmetrized quantities in \eq{DIS21} we can evaluate
the factors in the square brackets of \eqref{TMD10} as
\begin{align} \label{TMD11}
 \begin{aligned}
   \bigg[W_{unp}(p,b) \, D_{{\ul x} {\ul y}}[+\infty, b^-] &- W_{unp}(-\ul{p},b) \, D_{{\ul y} {\ul x}}[+\infty, b^-] \bigg] = \\
   &= 2 \bigg( W_{unp}^{OAM}(p,b) \, S_{{\ul x} {\ul y}}[+\infty, b^-]
   + W_{unp}^{symm}(p,b) \, i \, O_{{\ul x} {\ul y}}[+\infty, b^-]
   \bigg)
   \\
   \bigg[W_{trans}(p,b) \, D_{{\ul x} {\ul y}}[+\infty, b^-] &+ W_{trans}(-\ul{p},b) \, D_{{\ul y} {\ul x}}[+\infty, b^-]  \bigg] = \\
   &= 2 \bigg( W_{trans}^{symm}(p,b) \, S_{{\ul x} {\ul y}}[+\infty,
   b^-] + W_{trans}^{OAM}(p,b) \, i \, \, O_{{\ul x} {\ul y}}[+\infty,
   b^-] \bigg)
 \end{aligned}
\end{align}
giving
\begin{align} \label{TMD12} {\hat z} \cdot & (\ul{J}\times\ul{k}) \,
  f_{1T}^{\perp A} ({\bar x}, k_T) = M_A \int \frac{d p^+ \, d^2 p \,
    d b^-}{2(2\pi)^3} \, d^2x \, d^2 y \, \frac{d^2 k'}{(2\pi)^2} \,
  e^{ - i \, (\ul{k} - \ul{k'}) \cdot(\ul{x}-\ul{y})} \, \bigg\{
  f_1^N (x, |\ul{k}'- x \, \ul{p}|) \notag \\
  & \times \, A \, \left[ W_{unp}^{OAM} \left( p^+, \ul{p} ,b^-,
      \frac{{\ul x} + {\ul y}}{2} \right) \, S_{{\ul x} {\ul
        y}}[+\infty, b^-] + W_{unp}^{symm} \left( p^+ , \ul{p},b^-,
      \frac{{\ul x} + {\ul y}}{2} \right) \, i \, O_{{\ul x} {\ul
        y}}[+\infty, b^-] \right] \notag \\ & + \frac{1}{m_N} \, {\hat
    z} \cdot \left( \ul{S} \times (\ul{k}'- x \,
    \ul{p}) \right) \, f_{1T}^{\perp N} (x, |\ul{k}'- x \, \ul{p}|) \\
  & \times \, \left[ W_{trans}^{symm} \left( p^+, \ul{p} ,b^-,
      \frac{{\ul x} + {\ul y}}{2} \right) \, S_{{\ul x} {\ul
        y}}[+\infty, b^-] + W_{trans}^{OAM} \left( p^+, \ul{p} ,b^-,
      \frac{{\ul x} + {\ul y}}{2} \right) \, i \, O_{{\ul x} {\ul
        y}}[+\infty, b^-] \right] \bigg\}. \notag
\end{align}

Altogether, the symmetry arguments presented above allow us to
decompose the nuclear Sivers function $f_{1T}^{\perp A}$ into four
distinct channels with the right quantum numbers to generate the
$T$-odd asymmetry.  These four channels correspond to the negative
$T$-parity occurring in the nucleon distribution $W^{OAM}$, in the
quark Sivers function of the nucleon $f_{1T}^{\perp N}$, in the
antisymmetric ``odderon'' rescattering $i O_{xy}$, or in all three
simultaneously.

We now will neglect the odderon contributions in \eq{TMD12}. The way
to understand this approximation is as follows.  As shown in
\cite{Kovchegov:2012ga}, the preferred direction generated by
odderon-type rescattering couples to transverse gradients of the
nuclear profile function, $\ul{\nabla} T (\ul{b})$.  The length scale
over which these gradients become important is on the order of the
nuclear radius; these gradients are therefore $\ord{A^{-1/3}} \sim
\ord{\alpha_s^2}$ suppressed (in addition to an extra power of $\as$
entering the lowest-order odderon amplitude corresponding to the
triple-gluon exchange
\cite{Lukaszuk:1973nt,Nicolescu:1990ii,Ewerz:2003xi,Bartels:1999yt,Kovchegov:2003dm,Hatta:2005as,Kovner:2005qj,Jeon:2005cf,Kovchegov:2012ga})
and are beyond the precision of the quasi-classical formula
\eqref{TMD12}.

Neglecting the odderon channels in \eqref{TMD12} we arrive at
\begin{align} \label{TMD13}
  \begin{aligned} {\hat z} \cdot & (\ul{J}\times\ul{k}) \,
    f_{1T}^{\perp A} ({\bar x} , k_T) = M_A \, \int \frac{d p^+ \, d^2
      p \, d b^-}{2(2\pi)^3} \, d^2x \, d^2 y \, \frac{d^2
      k'}{(2\pi)^2} \, e^{ - i \, (\ul{k} - \ul{k'})
      \cdot(\ul{x}-\ul{y})}   \\
    & \times \, \bigg\{ A \, W_{unp}^{OAM} \left( p^+, \ul{p} ,b^-,
      \frac{{\ul x} + {\ul y}}{2} \right) \, f_1^N (x, |\ul{k}'- x \,
    \ul{p}|) \\ & + \frac{1}{m_N} \, {\hat z} \cdot \left( \ul{S}
      \times (\ul{k}'- x \, \ul{p}) \right) \, W_{trans}^{symm} \left(
      p^+, \ul{p} ,b^-, \frac{{\ul x} + {\ul y}}{2} \right) \,
    f_{1T}^{\perp N} (x, |\ul{k}'- x \, \ul{p}|) \bigg\} \, S_{{\ul x}
      {\ul y}}[+\infty, b^-] .
 \end{aligned}
\end{align}
Shifting the integration variable $\ul{k}' \to \ul{k}' + x \, \ul{p}$
we write
\begin{align} \label{TMD14}
  \begin{aligned} {\hat z} \cdot & (\ul{J}\times\ul{k}) \,
    f_{1T}^{\perp A} ({\bar x}, k_T) = M_A \, \int \frac{d p^+ \, d^2
      p \, d b^-}{2(2\pi)^3} \, d^2x \, d^2 y \, \frac{d^2
      k'}{(2\pi)^2} \, e^{ - i \, (\ul{k} - x \, \ul{p} - \ul{k'})
      \cdot(\ul{x}-\ul{y})}   \\
    & \times \, \bigg\{ A \ W_{unp}^{OAM} \left( p^+, \ul{p} ,b^-,
      \frac{{\ul x} + {\ul y}}{2} \right) \, f_1^N (x, k'_T) \\ & +
    \frac{1}{m_N} \, {\hat z} \cdot \left( \ul{S} \times \ul{k}'
    \right) \, W_{trans}^{symm} \left( p^+, \ul{p} ,b^-, \frac{{\ul x}
        + {\ul y}}{2} \right) \, f_{1T}^{\perp N} (x, k'_T) \bigg\} \,
    S_{{\ul x} {\ul y}}[+\infty, b^-] .
 \end{aligned}
\end{align}

\begin{figure}[ht]
\centering
\includegraphics[width= .4 \textwidth]{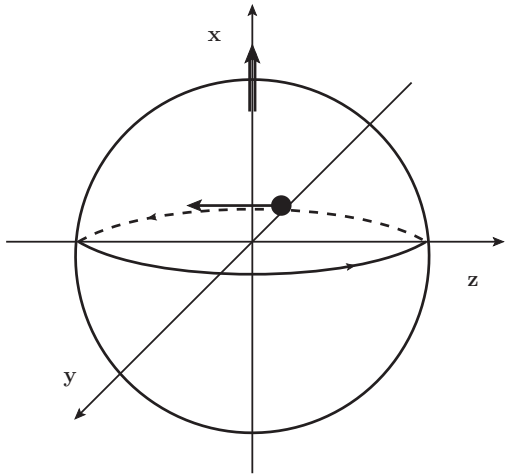}
\caption{This figure demonstrates our axes labeling convention and
  helps illustrate an example discussed in the text.}
\label{nucleus}
\end{figure}

To further simplify the obtained expression \eqref{TMD14} we need to
impose a constraint on the transverse momentum of the
nucleons. Consider the nucleus in its rest frame, as shown in
\fig{nucleus}.
The net OAM $\vec L$ of the transversely-polarized nucleus corresponds
to the rotation of the nucleus around the spin axis (the $x$-axis in
\fig{nucleus}). The rotational invariance around the $x$-axis implies
that the average magnitude of the rotational transverse momentum is
constant for a given distance from the $x$-axis and for fixed
$x$-coordinate. (In Appendix~\ref{B} we show that, as a consequence of
$PT$-symmetry, only rotational motion of the nucleons in the nucleus
rest frame is allowed.)

Consider a nucleon at the point ${\vec x} = (0,-R,0)$ in the $(x,y,z)$
coordinate system, as illustrated by the black circle in
\fig{nucleus}. Its 3-momentum is ${\vec p}_{rest} = (0,0,-p)$, where
$p$ denotes some typical rotational momentum of a nucleon. After a
longitudinal boost along the $z$-axis to the infinite-momentum frame
of \eqref{DIS1} we get the large light-cone component of the momentum
to be
\begin{align}
  p^+ = \frac{P^+}{M_A} \, \left( \sqrt{m_N^2 + p^2} - p \right).
\end{align}
The corresponding Bjorken-$x$ is (see \eq{DIS2})
\begin{align}
  \label{eq:mom2}
  1 \ge x = \frac{- q^+}{p^+} = x_A \, A \, \frac{m_N}{\sqrt{m_N^2 +
      p^2} - p },
\end{align}
where we have used $M_A = A \, m_N$. The $x \le 1$ constraint in
\eq{eq:mom2} (cf. \eq{DIS2}) gives
\begin{align}
  \label{eq:mom3}
  p \le m_N \, \frac{1 - x_A^2 \, A^2}{2 \, x_A \, A}.
\end{align}
Since $x_A \, A$ is not a small number, in fact $x_A \, A = \ord{1}$,
we conclude that $p \lesssim m_N$. Therefore, the magnitude of the
rotational momentum in the nuclear rest frame is bounded by $\sim m_N$
from above. The typical transverse momentum $p_T$ in \eq{TMD14}, being
boost-invariant, is also bounded by the nucleon mass from above, $p_T
\lesssim m_N$. Since we assume that $k_T$ is perturbatively large,
$k_T \gg \Lambda_{QCD} \sim m_N$, we do not consistently resum all
powers of $m_N/k_T$. (Saturation approach resums mainly
$A^{1/3}$-enhanced power corrections, that is, powers of
$Q_s^2/k_T^2$, but not powers of $\Lambda_{QCD}^2/k_T^2$.) 

The bound \eqref{eq:mom3} provides us with the condition on when the
SIDIS process on the nucleon highlighted in \fig{nucleus} can take
place. Violation of this bound would imply that SIDIS on that nucleon
is kinematically prohibited, and consequently SIDIS may take place
only on some of the other nucleons in the nucleus. While such
situation where the nucleus is spinning so fast that SIDIS is only
possible on a subset of its nucleons is highly unlikely in the real
physical experiments, this presents a theoretical example where the
Sivers function \eqref{TMD14} would, in fact, depend on the direction
of $\ul p$ and, hence, of spin $\ul {J}$, presumably through even
powers of $\ul{J} \cdot \ul{k}$. While such dependence is impossible
for spin-$1/2$ particles such as protons \cite{Bacchetta:1999kz}, it
has been considered for targets with different spin
\cite{Bacchetta:2000jk}; in our case it arises due to the classical
model at hand with the value of spin $J$ not restricted to $1/2$. To
avoid potential formal complications and unrealistic effects
associated with large rotational momentum, below we will assume that
$p_T \lesssim m_N$ such that the bound \eqref{eq:mom3} is
satisfied. Without such assumption, \eq{TMD14} would be our final
result for the Sivers function in the quasi-classical approximation.

We see that we have to limit the calculation to the lowest non-trivial
power of $p_T / k_T \sim m_N/k_T$ contributing to the Sivers
function. Expanding \eq{TMD14} in the powers of $\ul{p}$ to the lowest
non-trivial order, and remembering that $W^{OAM}$ is an odd function
of $\ul{p}$ we obtain
\begin{align} \label{TMD15}
  \begin{aligned} {\hat z} \cdot & (\ul{J}\times\ul{k}) \,
    f_{1T}^{\perp A} ({\bar x}, k_T) = M_A \,\int \frac{d p^+ \, d^2 p
      \, d b^-}{2(2\pi)^3} \, d^2x \, d^2 y \, \frac{d^2 k'}{(2\pi)^2}
    \, e^{ - i \, (\ul{k} - \ul{k'})
      \cdot(\ul{x}-\ul{y})}   \\
    & \times \, \bigg\{ i\, x \, \ul{p} \cdot(\ul{x}-\ul{y}) \, A \
    W_{unp}^{OAM} \left( p^+, \ul{p} ,b^-, \frac{{\ul x} + {\ul y}}{2}
    \right) \, f_1^N (x, k'_T) \\ & + \frac{1}{m_N} \, {\hat z} \cdot
    \left( \ul{S} \times \ul{k}' \right) \, W_{trans}^{symm} \left(
      p^+, \ul{p} ,b^-, \frac{{\ul x} + {\ul y}}{2} \right) \,
    f_{1T}^{\perp N} (x, k'_T) \bigg\} \, S_{{\ul x} {\ul y}}[+\infty,
    b^-] .
 \end{aligned}
\end{align}

\eq{TMD15} is our main formal result. It relates the Sivers function
of a nucleus to the quark TMD and quark Sivers function in a nucleon.
It shows that within the quasi-classical approximation, there are two
leading channels capable of generating the Sivers function of the
composite nucleus:
\begin{enumerate}
\item \ul{Orbital Angular Momentum (OAM) Channel}: an unpolarized
  nucleon in a transversely polarized nucleus with a preferred
  direction of transverse momentum generated by the OAM of the nucleus
  has a quark knocked out of its symmetric $f_1^N$ transverse momentum
  distribution which rescatters coherently on spectator nucleons. The
  multiple rescatterings bias the initial knockout process to happen
  near the ``back'' of the nucleus, where, due to OAM motion of the
  nucleons, the outgoing quark gets an asymmetric distribution of its
  transverse momentum, generating STSA. (See \fig{SIDIS_fig} or left
  panel of \fig{SIDIS_OAM_vs_Sivers} below.)
\item \ul{Transversity / Sivers Density Channel}: a polarized nucleon
  with its preferred transverse spin direction inherited from the
  nucleus has a quark knocked out of its Sivers $f_{1T}^{\perp N}$
  distribution which rescatters coherently on spectator nucleons. The
  single spin asymmetry is generated at the level of the ``first''
  nucleon, and, unlike the OAM channel, the presence of other nucleons
  is not essential for this channel (see \fig{SIDIS_OAM_vs_Sivers}).
\end{enumerate}

\begin{figure}[ht]
\centering
\includegraphics[width= .5 \textwidth]{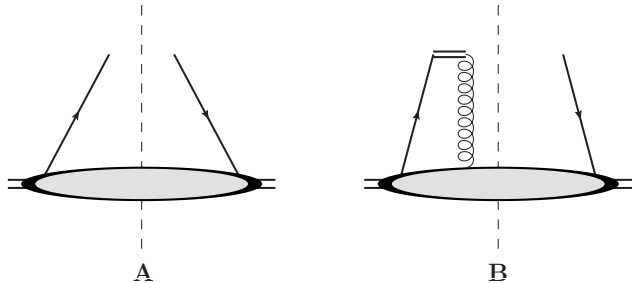}
\caption{Lowest-order diagrams for the quark TMD $f_1$ (panel A) and
  the Sivers function $f_{1T}^\perp$ (panel B). Vertical dashed line
  denotes the final state cut, while the double horizontal line in
  panel B denotes the Wilson line.}
\label{LO_TMDs}
\end{figure}

The OAM and transversity channels are depicted in
\fig{SIDIS_OAM_vs_Sivers} in terms of their space-time structure and
Feynman diagrams. The diagrams resummed in arriving at \eq{TMD15} are
the square of the graph shown in the left panel of
\fig{SIDIS_OAM_vs_Sivers} (OAM channel) and the diagram looking like
the interference between the two panels in \fig{SIDIS_OAM_vs_Sivers}
(transversity channel). The difference between the two channels
outlined above is in the first ``knockout'' interaction: the OAM
channel couples to quark TMD, while the transversity channel couples
to the nucleon Sivers function. At the lowest order in perturbation
theory the two functions are illustrated in \fig{LO_TMDs}: indeed
Sivers function shown in the panel B of \fig{LO_TMDs} requires at
least one more rescattering as compared to the quark TMD in panel A,
according to the conventional wisdom
\cite{Brodsky:2002cx,Collins:2002kn}.

\begin{figure}[h]
\centering
\includegraphics[width= 0.9 \textwidth]{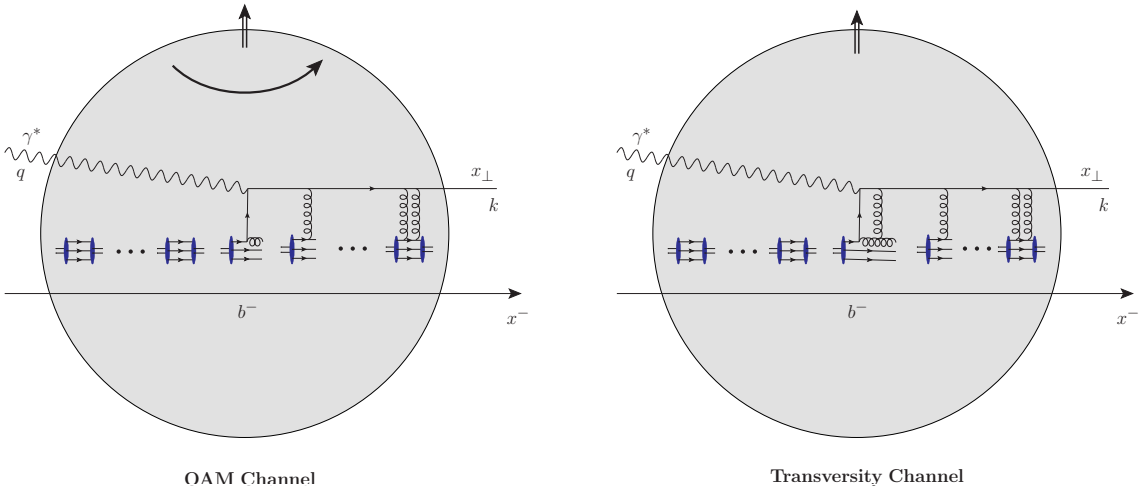}
\caption{Side-by-side comparison of the Feynman diagrams contribution
  to the OAM and Sivers density channels in the quasi-classical
  approximation (in the rest frame of the nucleus).}
\label{SIDIS_OAM_vs_Sivers}
\end{figure}

Note that, in the OAM channel, the unpolarized quark distribution
$f_1^N$ enters parametrically at $\ord{\alpha_s \, A^{1/3}}$ if
calculated at the lowest-order in the perturbation theory (see panel A
in \fig{LO_TMDs}), which is $\ord{\alpha_s^{-1}}$ in the saturation
power counting (where $\as^2 \, A^{1/3} \sim 1$).  In the transversity
channel, the nucleonic Sivers function $f_{1T}^{\perp N}$ enters at
$\ord{\alpha_s^2 \, A^{1/3}} = \ord{1}$ at the lowest order in
perturbation theory, because it requires an extra $\ord{\alpha_s}$
gluon to be exchanged with the same nucleon to obtain the necessary
lensing effect \cite{Brodsky:2002cx} (see panel B in
\fig{LO_TMDs}). The transversity channel is therefore $\ord{1}$ in the
saturation power counting and is subleading by $\ord{\alpha_s}$ to the
OAM channel in this sense.\footnote{We would like to point out that
  the coupling constant $\as$ in $f_1^N$ runs with some
  non-perturbative momentum scale, and is large, $\as = \as (\sim
  \Lambda_{QCD}^2)$; however, a simple application of the BLM
  \cite{Brodsky:1983gc} prescription to the calculation of
  \cite{Brodsky:2013oya} can show that in $f_{1T}^{\perp N} (x, k_T)$
  the two powers of the coupling run as $\as (k_T^2) \, \as (\sim
  \Lambda_{QCD}^2)$. While one of the couplings is also
  non-perturbatively large, the other one is perturbatively small for
  $k_T \gg \Lambda_{QCD}$, indicating suppression.} Indeed the
non-trivial transverse motion of nucleons due to OAM should be present
for the OAM channel to be non-zero: this channel is leading only if
there is an OAM. In our estimate here we have assumed that the net
spin of our ``nucleons'' scales linearly with the atomic number, $S
\sim A$; perhaps a more realistic (both for protons and nuclei) slower
growth of $S$ with $A$ would introduce extra $A$-suppression for the
transversity channel.

Despite the transversity channel being subleading, it is more dominant
than the $\ord{A^{-1/3}} \sim \ord{\alpha_s^2}$ corrections we
neglected when arriving at the quasi-classical formula \eqref{TMD15}
(again, for $S \sim A$). Order $\alpha_s^1$ quantum corrections to the
OAM channel also enter at the same order as the nucleonic Sivers
function and are also within the precision of the formalism.

An essential role is played by the rescattering factor $S_{{\ul x}
  {\ul y}}[+\infty, b^-]$ in the OAM channel. The Wigner function due
to orbital motion of nucleons around the axis of the transverse spin
is an odd function of the longitudinal coordinate $b_z$ in the rest
frame of the nucleus,
\begin{align}
  \label{eq:OAM_Wig}
  W_{unp}^{OAM} (p, \ul{b}, b_z) = - W_{unp}^{OAM} (p, \ul{b}, -b_z),
\end{align}
which follows simply from the fact that in the left panel of
\fig{SIDIS_OAM_vs_Sivers} we have as many nucleons moving outside the
page to the left of the nuclear center as there are nucleons moving
into the page to the right of the nuclear center. In Appendix~\ref{B}
we show how the result \eqref{eq:OAM_Wig} can be obtained by requiring
that our ``nucleus'' is $PT$-symmetric. The $b^-$-integral of the
first term in the curly brackets of \eq{TMD15} would have been zero,
if it was not for the $b^-$-dependent factor of $S_{{\ul x} {\ul
    y}}[+\infty, b^-]$. This multiple-rescattering factor approaches
$1$ for $b^-$ values near the ``back'' of the nucleus (right end of
the nucleus in \fig{SIDIS_OAM_vs_Sivers}) and is a monotonically
decreasing function of $b^-$. Due to this factor, different $b^-$
regions contribute differently to the integral, making it non-zero.
The region near the ``back'' of the nucleus dominates, which has a
physical interpretation that it is easier for the quark to escape the
nucleus if it is produced near the edge. Hence we arrive at the
interpretation of the SIDIS in the OAM channel outlined in the
Introduction: the quarks are produced predominantly toward the
``back'' of the nucleus, where the nucleons rotate preferentially into
the page (see left panel of \fig{SIDIS_OAM_vs_Sivers} or
\fig{SIDIS_fig}). Therefore, the quark has more transverse momentum
into the page than out of the page, which leads to STSA for the
produced quarks.

To complete \eq{TMD15} we need to construct an expression for the
nuclear spin ${\vec J} = {\vec L} + {\vec S}$. The OAM of the nucleons
in the nucleus from \fig{nucleus} in the nuclear rest frame is
\begin{align}
  \label{eq:OAM1}
  {\vec L} = A \int \frac{d^3 p \, d^3 b}{2 (2\pi)^3} \, W_{unp}
  \left( {\vec p}, \, {\vec b} \right) \ {\vec b} \times {\vec p} = A
  \int \frac{d^3 p \, d^3 b}{2 (2\pi)^3} \, W_{unp} \left( {\vec p},
    \, {\vec b} \right) \ {\hat x} \, (b_y \, p_z - b_z \, p_y),
\end{align}
where $d^3 p = d p_x \, d p_y \, d p_z$, $d^3 b = d b_x \, d b_y \, d
b_z$, and $W_{unp} \left( {\vec p}, \, {\vec b} \right)$ is the Wigner
distribution in the rest frame of the nucleus expressed in terms of
3-vectors ${\vec p} = (p_x, p_y, p_z)$ and ${\vec b} = (b_x, b_y,
b_z)$.

To boost this into the infinite momentum frame of \eqref{DIS1} we
define the Pauli-Lubanski vector of the nuclear spin
\begin{align}
  \label{eq:PL}
  W_\mu = - \frac{1}{2} \, \epsilon_{\mu\nu\rho\sigma} \, J^{\nu\rho}
  \, P^\sigma,
\end{align}
where $J_{\mu\nu} = L_{\mu\nu} + S_{\mu\nu}$ with $L_{\mu\nu}$ and
$S_{\mu\nu}$ the expectation values of the OAM and spin generators of
the Lorentz group in the nuclear state. The OAM generator is
\begin{align}
  \label{eq:generators}
  {\hat L}_{\mu\nu} = {\hat x}_\mu \, {\hat p}_\nu - {\hat x}_\nu \,
  {\hat p}_\mu
\end{align}
as usual, with the hat denoting operators. The nuclear OAM four-vector
is then defined by
\begin{align}
  \label{eq:OAM_4vector}
  L_\mu = - \frac{1}{2} \, \epsilon_{\mu\nu\rho\sigma} \, L^{\nu\rho}
  \, \frac{P^\sigma}{M_A}.
\end{align}
Note that ${\hat p}_\mu$ in \eq{eq:generators} are the momentum
operators of the nucleons, while $P^\sigma$ in Eqs.~\eqref{eq:PL} and
\eqref{eq:OAM_4vector} is the net momentum of the whole nucleus. In
the rest frame of the nucleus \eq{eq:OAM_4vector} gives $L_x = L_{yz}$
as expected (for $\epsilon_{0123} = +1$). The nuclear OAM four-vector
can then be written as
\begin{align}
  \label{eq:L}
  L_\mu = - \frac{1}{2} \, \epsilon_{\mu\nu\rho\sigma} \,
  \frac{P^\sigma}{M_A} \, A \, \int \frac{d p^+
      \, d^2 p \, d b^- \, d^2 b}{2 \, (2 \, \pi)^3} \, W_{unp} (p, b)
    \, (b^\nu \, p^\rho - b^\rho \, p^\nu)
\end{align}
in the infinite momentum frame of the nucleus.

Since boosts preserve transverse components of four-vectors, the boost
along the $\hat z$-axis of the nucleus in \fig{nucleus} would preserve
its OAM three-vector $\vec L$. Hence \eq{eq:OAM1} gives us the
transverse components of OAM in the infinite momentum frame as
well. We thus write
\begin{align}
  \label{eq:J} {\vec J} = {\hat x} \left[ S + A \int \frac{d^3 p \,
      d^3 b}{2 (2\pi)^3} \, W_{unp} \left( {\vec p}, \, {\vec b}
    \right) \ {\hat x} \, (b_y \, p_z - b_z \, p_y) \right],
\end{align}
where the integration over $p$ and $b$ needs to be carried out in the
nucleus rest frame.  

Combining Eqs.~\eqref{TMD15} with \eqref{eq:J} allows one to extract
the Sivers function $f_{1T}^{\perp A}$ of the nucleus.


\subsection{Comparison of the OAM and Transversity Channels in the SIDIS Sivers Function}
\label{toy_model}

We will now illustrate the properties of the Sivers function
\eqref{TMD15} by studying a specific simplified example. Consider the
model of the nucleus as a non-relativistic rigid rotator, with the
rotational momentum in its rest frame being much smaller than the
nucleon mass, $p_T \ll m_N$. The corresponding classical Wigner
distribution is (cf. \eq{Wcl})
\begin{align}
\label{Wcl_oam}
W_{unp} (p,b) \approx \frac{2 \, (2 \, \pi)^3}{A} \, \rho ({\ul b},
b^-) \, \delta^2 \left(\ul{p} - \hat{y} \, p_{max} (b_x) \,
  \frac{b^-}{R^- (b_x)} \right) \, \delta \left( p^+ - \frac{P^+}{A}
\right),
\end{align}
where $2 R^- (b_x)$ is the extent of the nucleus in the
$b^-$-direction at $\ul{b} = (b_x, 0)$ (with $R^-(b_x) = \sqrt{R^2 -
  b_x^2} \ M_A/P^+$), and $p_{max} (b_x) = p_{max} \, \sqrt{R^2 -
  b_x^2}/R$ is the maximum value of the rotational momentum at a given
$b_x$.  In writing down the distribution \eqref{Wcl_oam} we have
neglected possible longitudinal rotational motion of the nucleons,
which is justified in the $p_T \ll m_N$ limit.  We also assume that
a fraction $\beta$ of the nucleons in the nucleus are polarized in the 
$+{\hat x}$-direction, such that their net spin is $S = \beta A/2$ and (see
\eq{TMD8})
\begin{align} 
\label{rr1b} 
W_{trans} (p,b) = \beta W_{unp}(p,b).
\end{align}

Substituting Eqs.~\eqref{Wcl_oam} and \eqref{rr1b} into \eq{TMD15} and
integrating over $p^+$ and $\ul p$ yields
\begin{align} 
  \label{SiversA} & J \, k_y \, f_{1T}^{\perp A} ({\bar x} , k_T) =
  M_A \int d b^- \, d^2 x \, d^2 y \, \rho \left( \frac{{\ul x} + {\ul
        y}}{2} , b^-\right) \, \frac{d^2 k'}{(2\pi)^2} \, e^{ - i \,
    (\ul{k} - \ul{k'}) \cdot ({\ul x} - {\ul y})} 
    \bigg\{ i \, {\bar x} \, \notag \\
  & \hspace{-0.2cm} \times p_{max} \left( \frac{({\ul x} +
      {\ul y})_x}{2} \right) \, \frac{b^-}{R^- ((\tfrac{x+y_x}{2})_x)} 
      ({\ul x} - {\ul
    y})_y \, f_1^N ({\bar x} , k'_T) + \frac{\beta}{2 \, m_N} \, k'_y \,
  f_{1T}^{\perp N} ({\bar x}, k'_T ) \bigg\} \, S_{{\ul x} {\ul
      y}}[+\infty, b^-],
\end{align}
where we also replaced ${\vec J}$ by ${\hat x} \, J$ and ${\vec S}$ by
${\hat x} \, (\beta A/2)$.

To further simplify \eq{SiversA} we need to make specific assumptions
about the form of $f_1^N$ and $f_{1T}^{\perp N}$. Inspired by the
lowest-order expressions for both quantities
\cite{Itakura:2003jp,Boer:2002ju,Brodsky:2013oya} we write
\begin{align}
  \label{eq:fs}
  f_1^N (x, k_T) = \frac{\as \, C_1}{k_T^2}, \ \ \ f_{1T}^{\perp N}
  (x, k_T) = \frac{\as^2 \, m_N^2 \, C_2}{k_T^4} \, \ln
  \frac{k_T^2}{\Lambda^2},
\end{align}
where $C_1$ and $C_2$ are some $x$-dependent functions and $\Lambda$
is an infrared cutoff. Inserting \eq{eq:fs} into \eq{SiversA} and
integrating over $k'_T$ yields
\begin{align} 
  \label{SiversA2} & J \, k_y \, f_{1T}^{\perp A} ({\bar x} , k_T) =
  \frac{\as \, M_A}{2\pi} \int d b^- \, d^2x \, d^2 y \, \rho
  \left(\frac{{\ul x} + {\ul y}}{2} , b^-\right) \, e^{ - i \,
    \ul{k} \cdot(\ul{x}-\ul{y})} \, (\ul{x}-\ul{y})_y   
    \bigg\{ i \, {\bar x} 
  \notag \\
  & \times \, p_{max}\left( \frac{({\ul x} +
      {\ul y})_x}{2} \right) \, \frac{b^- \, C_1}{R^- ((\tfrac{x+y}{2})_x)} \,
  \ln\frac{1}{|\ul{x} -\ul{y}| \Lambda} + \frac{i \, \as \, m_N \, \beta \,
    C_2}{4} \, \ln^2 \frac{1}{|\ul{x} -\ul{y}| \Lambda} \bigg\} \,
  S_{{\ul x} {\ul y}}[+\infty, b^-].
\end{align}

In the classical MV/GM approximation the (symmetric part of the)
dipole scattering matrix is \cite{Mueller:1989st}
\begin{align} 
  \label{Wlines4} S_{{\ul x} {\ul y}}[+\infty, b^-] = \exp
  \left[-\frac{1}{4} |\ul{x}- \ul{y}|^2 \, Q_s^2 \left(\frac{{\ul x} +
        {\ul y}}{2} \right) \, \left(\frac{R^- (\ul{b}) - b^-}{2 R^-
        (\ul{b})}\right) \, \ln\frac{1}{|\ul{x} -\ul{y}| \Lambda}
  \right],
\end{align}
where $R^-(\ul{b}) = \sqrt{R^2 - \ul{b}^2} \ M_A/P^+$ and the quark
saturation scale is
\begin{align} \label{Wlines6}
 Q_s^2(\ul{b})~=~4\pi \, \as^2 \, \tfrac{C_F}{N_c} \, T(\ul{b})
\end{align}
with the nuclear profile function
\begin{align}
  \label{eq:nuc_prof}
  T(\ul{b}) = \int d b^- \, \rho \left( \ul{b} , b^-\right).
\end{align}
As usual $N_c$ is the number of colors and $C_F = (N_c^2 -1)/2 N_c$ is
the Casimir operator of SU$(N_c)$ in the fundamental representation.
In arriving at \eq{Wlines4} we assumed that the nuclear density is
constant within the nucleus, such that
\begin{align}
  \label{eq:dens}
  \rho \left( \ul{b} , b^-\right) = \frac{\theta (R^- (\ul{b}) -
    |b^-|)}{2 R^- (\ul{b})} \, T(\ul{b}).
\end{align}

Employing \eq{Wlines4} along with Eqs.~\eqref{eq:dens} and
\eqref{Wlines6}, and neglecting all logarithms $\ln (1/|\ul{x}
-\ul{y}| \Lambda)$ (which is justified as long as $k_T$ is not too
much larger than $Q_s$ \cite{Kovchegov:1998bi}) we can integrate
\eq{SiversA2} over $b^-$ and $\ul{x} -\ul{y}$ obtaining
\begin{align} 
  \label{SiversA3} f_{1T}^{\perp A} ({\bar x} , k_T) = \frac{M_A \,
    N_c}{4\pi \, \as \, J \, C_F} \, \frac{1}{k_T^2} \int d^2b \,
  \bigg\{ & 4 \, {\bar x} \, p_{max} (\ul{b}) \, C_1 \, \left[
    e^{-k_T^2/Q_s^2 ({\ul b})} + 2 \, \frac{k_T^2}{Q_s^2 ({\ul b})} \,
    Ei \left(- \frac{k_T^2}{Q_s^2 ({\ul b})}\right)\right]  \notag \\
  & + \as \, \beta \, m_N \, C_2 \, e^{-k_T^2/Q_s^2 ({\ul b})} \bigg\},
\end{align}
where now $\ul{b} = (\ul{x} + \ul{y})/2$ and $p_{max} (\ul{b}) =
p_{max} \, \sqrt{R^2 - \ul{b}^2}/R$. The $\ul b$-integral appears to
be rather hard to perform for a realistic spherical nucleus: we leave
expression \eqref{SiversA3} in its present form.

To obtain a final expression for the Sivers function we need to
determine the spin of the nucleus $J$. For a rigid rotator spinning
around the $\hat x$-axis with the maximum nucleon momentum $p_{max}$
we readily get
\begin{align}
  \label{eq:Lrigid}
  L = \frac{4}{5} \, A \, p_{max} \, R
\end{align}
in the nuclear rest frame. Using this in \eq{eq:J} along with $S =
\beta \, A/2$ we obtain
\begin{align}
  \label{eq:Jmod}
  J = \beta \frac{A}{2} + \frac{4}{5} \, A \, p_{max} \, R.
\end{align}
Inserting \eq{eq:Jmod} into \eq{SiversA3} gives
\begin{align} 
  \label{SiversA4} f_{1T}^{\perp A} & ({\bar x} , k_T) = \frac{m_N \,
    N_c}{2\pi \, \as \, C_F} \, \frac{1}{\beta + \tfrac{8}{5} \, p_{max} \,
    R} \, \frac{1}{k_T^2} \notag \\
  \times \int & d^2b \, \bigg\{ 4 \, {\bar x} \, p_{max} (\ul{b}) \,
  C_1 \, \left[ e^{-k_T^2/Q_s^2 ({\ul b})} + 2 \, \frac{k_T^2}{Q_s^2
      ({\ul b})} \, Ei \left(- \frac{k_T^2}{Q_s^2 ({\ul
          b})}\right)\right] + \as \beta \, m_N \, C_2 \, e^{-k_T^2/Q_s^2
    ({\ul b})} \bigg\}.
\end{align}
\eq{SiversA4} is our final expression for the Sivers function of a
nucleus in the quasi-classical approximation with the rigid rotator
model for the nucleus and $k_T$ not too much larger than
$Q_s$. Analyzing this expression we see that the OAM term (the first
term in the curly brackets) does change sign as a function of $k_T$,
while the Sivers density term (the second term in the curly brackets
of \eq{SiversA4}) is positive-definite. Still the first term in the
curly brackets is positive for most of the $k_T$ domain, in agreement
with the qualitative argument in the Introduction corresponding to
quarks being produced preferentially into the page in
\fig{SIDIS_fig}. 

To study the $k_T \gg Q_s$ case we have to return to \eq{SiversA2}:
this time we do not neglect the logarithms. Large $k_T$ limit implies
that $|\ul{x} - \ul{y}|$ is small, and we need to expand the
exponential in \eq{Wlines4} to the lowest non-trivial (contributing)
order in each term in \eq{SiversA2}. For the Sivers density term this
corresponds to replacing the exponent by $1$. The remaining evaluation
is easier to carry out in \eq{SiversA}, which yields an intuitively
clear formula
\begin{align}
  \label{eq:transv}
  J \, f_{1T}^{\perp A} ({\bar x}, k_T) \big|_{transversity \ channel, \ k_T
    \gg Q_s} = A \, S \, f_{1T}^{\perp N} ({\bar x}, k_T).
\end{align}
In the first term in the curly brackets of \eq{SiversA2} we need to
expand the exponential in $S_{{\ul x} {\ul y}}[+\infty, b^-]$ one step
further, obtaining after some straightforward algebra for the whole
SIDIS Sivers function
\begin{align}
  \label{eq:Sivers_highk}
  & f_{1T}^{\perp A} ({\bar x}, k_T) \big|_{k_T \gg Q_s} = \frac{S}{J} \left[
    - \frac{4 \, \as \, m_N \, {\bar x} \, C_1}{3 \beta \, k_T^6} \, \ln
    \frac{k_T^2}{\Lambda^2} \, \int d^2 b \, T(\ul{b}) \, p_{max}
    (\ul{b}) \, Q_s^2 (\ul{b}) + A \,
    f_{1T}^{\perp N} ({\bar x}, k_T) \right] \notag \\
  = & \frac{\beta}{\beta+ \tfrac{8}{5} \, p_{max} \, R} \left[ - \frac{4 \,
      \as \, m_N \, {\bar x} \, C_1}{3 \beta \, k_T^6} \ln \frac{k_T^2}{\Lambda^2}
    \int d^2 b \ T(\ul{b}) \, p_{max} (\ul{b}) \, Q_s^2 (\ul{b}) +
    \frac{A \, \as^2 \, m_N^2 \, C_2}{k_T^4} \, \ln
    \frac{k_T^2}{\Lambda^2} \right].
\end{align}
Since 
\begin{align}
  \label{eq:nucl_prof}
  \int d^2 b \, T(\ul{b}) = A
\end{align}
we see that the OAM channel contribution in \eq{eq:Sivers_highk} (the
first term) is proportional to $A \, \as \, m_N \, p_T \,
Q_s^2/k_T^6$, while the transversity channel contribution (the second
term) scales as $A \, \as^2 \, m_N^2 / k_T^4$. (Note that $x =
\ord{1}$, such that powers of $x$ do not generate suppression.)
Assuming that $p_T \approx m_N$ (see the discussion following
\eq{eq:mom3}) we observe that the ratio of the OAM to transversity
channel contributions is $\sim Q_s^2/(\as \, k_T^2)$. (Note that for
$p_T \approx m_N$ the prefactor of \eq{eq:Sivers_highk} gives a factor
$\sim 1/(m_N \, R) \approx A^{-1/3}$ multiplying both terms, but not
affecting their ratio.) We conclude that the OAM channel dominates for
\begin{align}
  \label{eq:bound}
  k_T < \frac{Q_s}{\sqrt{\as}},
\end{align}
that is both inside the saturation region, and in a sector of phase
space outside that region. For $k_T > Q_s/\sqrt{\as}$ the transversity
channel dominates, mapping onto the expected perturbative QCD result
\eqref{eq:transv}.

While the main aim of this calculation is to model a nucleon at high
energies, a few comments are in order about the application of this
rigid rotator toy model to a realistic nucleus.  Certainly a classical
rigid rotator is a poor model for a real nucleus; a better approach
would be to use our general result \eqref{TMD15} with the Wigner
distribution $W(p,b)$ given by the realistic single-particle wave
functions taken from nuclear structure calculations.  In such
realistic cases, the total angular momentum $J$ of the nucleus is
typically small, and the fraction $\beta$ that comes from the
nucleons' spins is also small due to nucleon spin pairing.  If one
were to approximate a real nucleus with this rigid rotator toy model,
appropriately small $J$ and $\beta$ would need to be used in
\eqref{SiversA4} and \eqref{eq:Sivers_highk}.  The smallness of the
total OAM $J$ does not affect the Sivers function $f_{1T}^{\bot A}$
because the magnitude is contained in the prefactor ${\hat z} \cdot
(\ul{J} \times \ul{k})$ and cancels in the $S/J$ ratio.  The smallness
of the spin contribution $\beta \sim \mathcal{O}(1/A)$, however, would
suppress the transversity channel and ensure the dominance of the OAM
term.  But regardless of its applicability to a real nucleus, the
rigid rotator toy model illustrates the ability of this formalism to
capture the interplay of spin and angular momentum in a dense system
at high energy.


\section{Drell-Yan Process}
\label{sec:DY}

We now wish to perform a similar analysis for the Drell-Yan process
$\overline{q} + A^\uparrow \rightarrow \gamma^* + X \rightarrow \ell^+
\ell^- + X$, where the antiquark from the unpolarized hadron scatters
on the transversely polarized hadron/nucleus, producing a space-like
photon which later decays into a di-lepton pair. The annihilation
sub-process $\overline{q} + q^\uparrow \rightarrow \gamma^* + X$ is
related to the SIDIS process by time reversal, which leads to the
famous prediction \cite{Collins:2002kn} that the Sivers functions
entering observables in the two processes should have equal magnitudes
and opposite signs.

\begin{figure}[ht]
\centering
\includegraphics[width = 0.6 \textwidth]{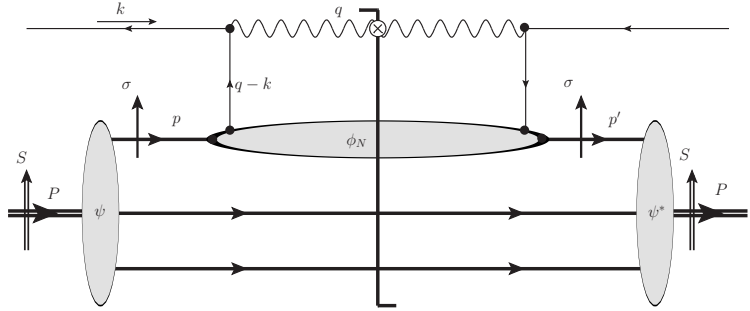}
\caption{
  \label{fig:DY1} Lowest-order DY process in the usual $\alpha_s$
  power-counting.  An antiquark from a projectile hadron annihilates
  with a quark from a nucleon in the target nucleus, producing a
  highly virtual photon which then decays into a di-lepton pair (not
  shown).  }
\end{figure}

The lowest-order Drell-Yan annihilation process is shown in
Fig.~\ref{fig:DY1}, without including initial-state rescattering of
the antiquark on nuclear spectators. Labeling the momenta as in
Fig.~\ref{fig:DY1} and following along the same lines as for SIDIS, we
can write the kinematics in the $\overline{q}+ A^\uparrow$
center-of-mass frame as
\begin{align} \label{DY1}
 \begin{aligned}
   P^\mu &= \big(P^+ , M_A^2 / P^+ , \ul{0}\big) \\
   p^\mu &= \big(p^+ , (p_T^2 + m_N^2) / p^+ , \ul{p}\big) \\
   k^\mu &= \big(\tfrac{m_q^2}{Q^2} \, q^+ , k^- \approx q^-, \ul{0}\big) \\
   q^\mu &= \big(q^+ , q^- \approx Q^2/q^+ , \ul{q}\big),
 \end{aligned}
\end{align}
where 
\begin{align} \label{DY2}
 \begin{aligned}
   \hat{s} &\equiv (p+k)^2 \approx p^+ q^- \\
   x &\equiv \frac{Q^2}{2 p \cdot q} \approx \frac{Q^2}{\hat{s}}
   \approx \frac{q^+}{p^+}.
 \end{aligned}
\end{align}
As with SIDIS, we are working in the kinematic limit $s_A = (P+k)^2
\gg \hat{s}, Q^2 \gg \bot^2$, with $\alpha \equiv p^+ / P^+ \approx
s_A / \hat{s} \sim \ord{1/A}$.  Again, we can compare the coherence
lengths $\ell_k^- \sim 1 / k^+$ of the antiquark and $\ell_\gamma^-
\sim 1/q^+$ of
\begin{align} \label{DY3}
 \begin{aligned}
   \frac{\ell_k^-}{L^-} &\sim \frac{1}{x}
   \left(\frac{Q^2}{m_q^2}\right) \frac{1}{\alpha \, M_A \, R}
   \sim \ord{\frac{Q^2}{m_q^2} \, A^{-1/3}} \gg 1 \\
   \frac{\ell_\gamma^-}{L^-} &\sim \frac{1}{x} \frac{1}{\alpha \, M_A
     \, R} \sim \ord{A^{-1/3}} \ll 1.
 \end{aligned}
\end{align}
Analogous to SIDIS, this shows that the coherence length of the
incoming antiquark is large; in fact it would be infinite if we
dropped the quark mass $m_q$ as we have elsewhere in the calculation.
We conclude that the long-lived antiquark is able to rescatter off of
many nucleons before it finally annihilates a quark. The annihilation
occurs locally, as indicated by the short coherence length of the
virtual photon, and thereafter the produced photon / dilepton system
does not rescatter hadronically. This again motivates the resummation
of these initial state rescatterings into a Wilson line dipole trace.


\subsection{Quasi-Classical Sivers Function in DY}

\begin{figure}[ht]
\centering
\includegraphics[width= .5 \textwidth]{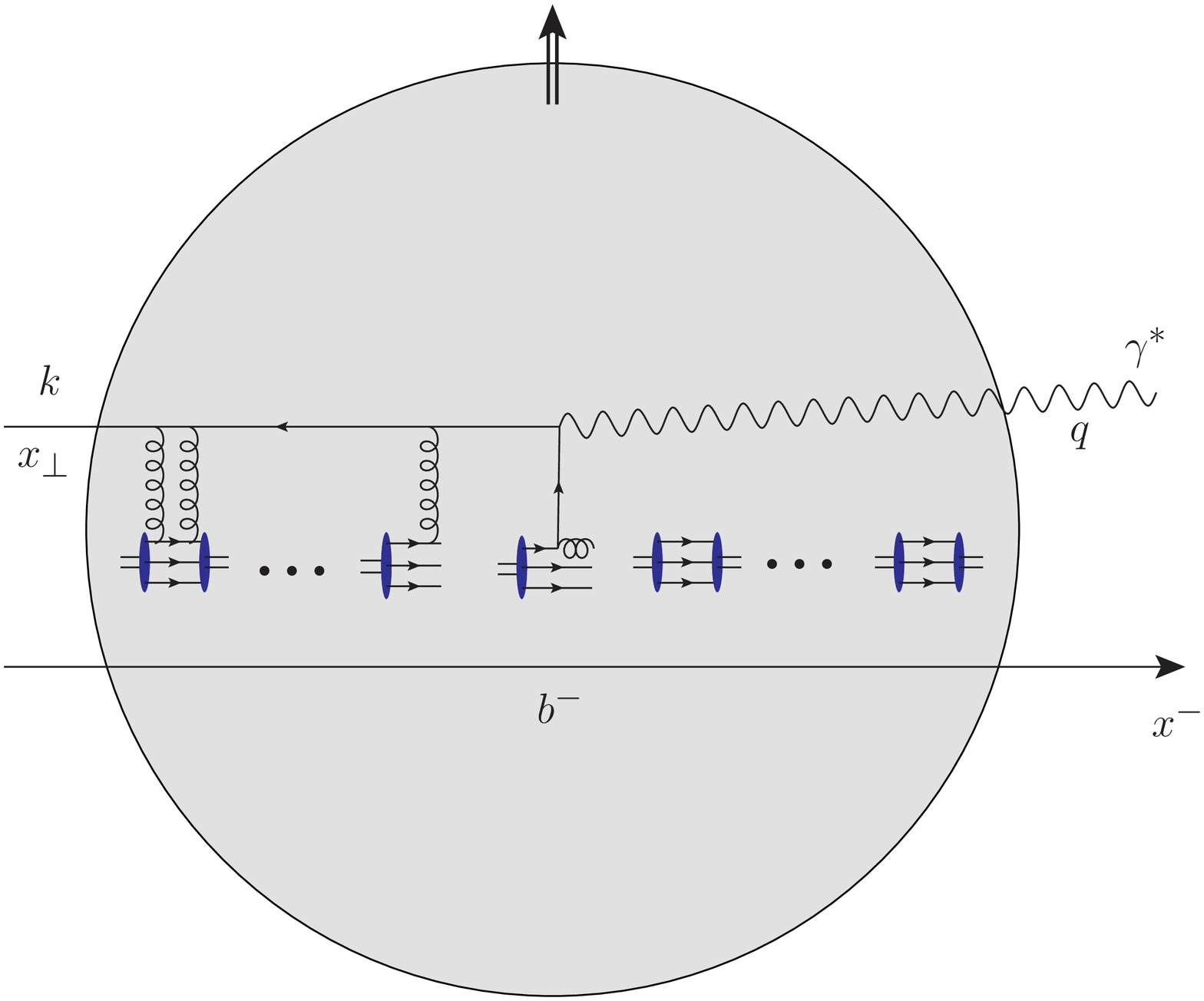}
\caption{Space-time structure of the quasi-classical DY process in the
  rest frame of the nucleus, overlaid with one of the corresponding
  Feynman diagrams. The shaded circle is the transversely polarized
  nucleus, with the vertical double arrow denoting the spin
  direction.}
\label{DY_ampl}
\end{figure}

The entire Drell-Yan (DY) process in the quasi-classical approximation
is shown in \fig{DY_ampl} at the level of the scattering amplitude:
the incoming antiquark coherently scatters on the nucleon in the
transversely polarized nucleons, until the last interaction in which
the virtual photon is produced, which later generated the di-lepton
pair.\footnote{Just like for SIDIS, in small-$x$ physics the DY
  process is dominated by the ${\bar q} + A \to \gamma^* + {\bar q} +
  A$ scattering \cite{Kopeliovich:2001hf}, which is
  $\ord{\as}$-subleading compared to the diagram in \fig{DY_ampl}:
  since in our calculation $x = \ord{1}$, we will neglect the ${\bar
    q} + A \to \gamma^* + {\bar q} + A$ process here as an $\ord{\as}$
  correction.}

\begin{figure}[ht]
\centering
\includegraphics[width = 0.7 \textwidth]{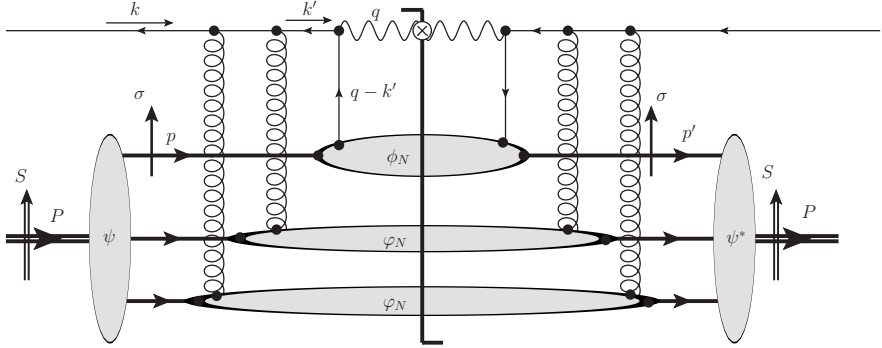}
\caption{
  \label{fig:DY2} Decomposition of the nuclear quark distribution
  $\Phi_A$ probed by the DY process into mean-field wave functions
  $\psi, \psi^*$ of nucleons and the quark and gluon distributions
  $\phi_N$ and $\varphi_N$ of the nucleons.  }
\end{figure}

By analogy with \eq{TMD1} in SIDIS we write the following relation for
the quark correlators in DY,
\begin{align} \label{DYTMD1}
 \begin{aligned}
   \Tr[\Phi_A({\bar x}, \ul{q}; P, J) \gamma^+ ] = A \int & \frac{d
     p^+ \, d^2 p \, d b^-}{2(2\pi)^3} \int \frac{d^2 k' \, d^2 x \, d^2
     y}{(2\pi)^2} \, e^{i \ul{k'} \cdot (\ul{x} - \ul{y})} \,
   \sum_\sigma W_N^\sigma \left(p, b^-, \frac{\ul{x} +
       \ul{y}}{2} \right) \\
   &\times \Tr[\phi_N (x , \ul{q} -\ul{k}'-x \, \ul{p}); p, \sigma)
   \gamma^+] \, D_{{\ul{y} \, \ul{x}}} [b^-, -\infty],
 \end{aligned}
\end{align}
where 
\begin{align}
 \label{dipole_def_DY}
 D_{{\ul y} \, {\ul x}} [b^-, -\infty] = \left\langle \frac{1}{N_c} \,
   \mbox{Tr} \left[ V_{\ul y} [b^-, -\infty] \, V^\dagger_{\ul x}
     [b^-, -\infty] \right] \right\rangle
\end{align}
and the quark correlators are defined by equations similar to
\eqref{TMD2} and \eqref{TMD3}, but now using a different gauge link
\eqref{eq:U_DY}:
\begin{align}
  \label{DYcorr2}
  \Phi_{ij}^A ({\bar x}, {\ul k}; P, J) \equiv \int \frac{d x^- \, d^2
    x_\perp}{2(2 \, \pi)^3} \, e^{i \, \left(\frac{1}{2} \, {\bar x} \,
      P^+ \, x^- - {\ul x} \cdot {\ul k} \right)} \, \langle A; P, J|
  {\bar \psi}_j (0) \, {\cal U}^{DY} \, \psi_i (x^+=0, x^-, {\ul x}) |
  A; P, J \rangle ,
\end{align}
\begin{align} 
  \label{DYcorr3} 
  \phi_{ij}^N (x, {\ul k}; p, \sigma) \equiv \int \frac{d x^- \, d^2
    x_\perp}{2(2 \, \pi)^3} \, e^{i \, \left(\frac{1}{2} \, x \, p^+ \,
      x^- - {\ul x} \cdot {\ul k} \right)} \, \langle N; p, \sigma |
  {\bar \psi}_j (0) \, {\cal U}^{DY} \, \psi_i (x^+=0, x^-, {\ul x}) |
  N; p, \sigma \rangle.
\end{align}
Here ${\bar x} = A \, q^+/P^+$. \eq{DYTMD1} is illustrated in
\fig{fig:DY2}. The main difference compared to \eq{TMD1} is that now
$\ul{k} =0$ and $\ul{q} \neq 0$.

Projecting out the DY Sivers function of the nucleus $f_{1T}^{\perp
  A}$ using \eqref{TMD5} gives
\begin{align} 
\label{DY_TMD2}
   {\hat z} \cdot (\ul{J}\times\ul{q}) \, f_{1T}^{\perp A} ({\bar x} ,
   q_T) = & \, \frac{M_A \, A}{4} \int \frac{d p^+ \, d^2 p \, d
     b^-}{2(2\pi)^3} \int \frac{d^2 k' \, d^2 x \, d^2 y}{(2\pi)^2} \,
   e^{i \ul{k'} \cdot (\ul{x} - \ul{y})} \, \sum_\sigma W_N^\sigma
   \left(p, b^-, \frac{\ul{x} +
       \ul{y}}{2} \right)  \notag \\
   &\times \Tr[\phi_N (x , \ul{q} -\ul{k}'-x \, \ul{p}); p, \sigma)
   \gamma^+] \, D_{{\ul{y} \, \ul{x}}} [b^-, -\infty] - (\ul{q}
   \rightarrow -\ul{q}).
\end{align}
With the help of \eq{TMD4B} we write
\begin{align} 
\label{DY_TMD3}
   {\hat z} \cdot & (\ul{J}\times\ul{q}) \, f_{1T}^{\perp A} ( {\bar
     x} , q_T) = \, \frac{M_A \, A}{2} \int \frac{d p^+ \, d^2 p \, d
     b^-}{2(2\pi)^3} \int \frac{d^2 k' \, d^2 x \, d^2 y}{(2\pi)^2} \,
   e^{i \ul{k'} \cdot (\ul{x} - \ul{y})} \, \sum_\sigma W_N^\sigma
   \left(p, b^-, \frac{\ul{x} +
       \ul{y}}{2} \right) \notag  \\
   & \times \bigg[ f_1^N (x , |\ul{q} -\ul{k}' - x \, \ul{p}|_T) +
   \frac{1}{m_N} {\hat z} \cdot \left(\ul{\sigma} \times (\ul{q}
     -\ul{k}'-x \, \ul{p})\right) \, f_{1T}^{\perp N} (x , |\ul{q}
   -\ul{k}'- x \ul{p}|_T) \bigg] \notag  \\
   & \times \, D_{{\ul{y} \, \ul{x}}} [b^-, -\infty] - (\ul{q}
   \rightarrow -\ul{q}).
\end{align}
Performing the spin sums \eqref{TMD8} gives
\begin{align} 
  \label{DY_TMD4} {\hat z} \cdot & (\ul{J}\times\ul{q}) \,
  f_{1T}^{\perp A} ( {\bar x} , q_T) = \, \frac{M_A}{2} \int \frac{d
    p^+ \, d^2 p \, d b^-}{2(2\pi)^3} \int \frac{d^2 k' \, d^2 x \,
    d^2 y}{(2\pi)^2} \, e^{i \ul{k'} \cdot (\ul{x} - \ul{y})} \,
  \bigg[ A \, W_{unp} \left(p, b^-, \frac{\ul{x} +
      \ul{y}}{2} \right) \notag  \\
  & \times \, f_1^N (x , |\ul{q} -\ul{k}' - x \, \ul{p}|_T) +
  W_{trans} \left(p, b^-, \frac{\ul{x} + \ul{y}}{2} \right) \,
  \frac{1}{m_N} {\hat z} \cdot \left(\ul{S} \times (\ul{q}
    -\ul{k}'-x \, \ul{p})\right) \notag  \\
  & \times \, f_{1T}^{\perp N} (x , |\ul{q} -\ul{k}'- x \ul{p}|_T)
  \bigg] \, D_{{\ul{y} \, \ul{x}}} [b^-, -\infty] - (\ul{q}
  \rightarrow -\ul{q}).
\end{align}
In the terms being subtracted in \eq{DY_TMD4} with $(\ul{q}
\rightarrow -\ul{q})$, we can also reverse the dummy integration
variables $\ul{k'} \rightarrow -\ul{k'}$, $\ul{p} \rightarrow -
\ul{p}$, and $\ul{x} \leftrightarrow \ul{y}$.  This leaves the Fourier
factor and the distributions $f_1^N$, $f_{1T}^{\perp N}$ invariant,
giving
\begin{align} 
  \label{DY_TMD5} {\hat z} \cdot & (\ul{J}\times\ul{q}) \,
  f_{1T}^{\perp A} ( {\bar x} , q_T) = \, \frac{M_A}{2} \int \frac{d
    p^+ \, d^2 p \, d b^-}{2(2\pi)^3} \int \frac{d^2 k' \, d^2 x \,
    d^2 y}{(2\pi)^2} \, e^{i \ul{k'} \cdot (\ul{x} - \ul{y})} \,
  \bigg\{ f_1^N (x , |\ul{q} -\ul{k}' - x \, \ul{p}|_T) \notag \\
  & \times \, A \, \bigg[ W_{unp} \left(p^+, \ul{p} , b^-,
    \frac{\ul{x} + \ul{y}}{2} \right) \, D_{{\ul{y} \, \ul{x}}} [b^-,
  -\infty] - W_{unp} \left(p^+, - \ul{p} , b^-, \frac{\ul{x} +
      \ul{y}}{2} \right) \, D_{{\ul{x} \, \ul{y}}} [b^-, -\infty]
  \bigg] \notag \\ & + \frac{1}{m_N} {\hat z} \cdot \left(\ul{S}
    \times (\ul{q} -\ul{k}'-x \, \ul{p})\right) \, f_{1T}^{\perp N} (x
  , |\ul{q} -\ul{k}'- x \ul{p}|_T) \\
  & \times \bigg[ W_{trans} \left(p^+, \ul{p}, b^-, \frac{\ul{x} +
      \ul{y}}{2} \right) \, D_{{\ul{y} \, \ul{x}}} [b^-, -\infty] +
  W_{trans} \left(p^+, - \ul{p}, b^-, \frac{\ul{x} + \ul{y}}{2}
  \right) \, D_{{\ul{x} \, \ul{y}}} [b^-, -\infty] \bigg]
  \bigg\}. \notag
\end{align}
We recognize the factors in brackets from the SIDIS case
\eqref{TMD11}, rewriting \eqref{DY_TMD5} as
\begin{align} 
  \label{DY_TMD6} {\hat z} \cdot & (\ul{J}\times\ul{q}) \,
  f_{1T}^{\perp A} ( {\bar x} , q_T) = \, M_A \int \frac{d
    p^+ \, d^2 p \, d b^-}{2(2\pi)^3} \int \frac{d^2 k' \, d^2 x \,
    d^2 y}{(2\pi)^2} \, e^{i \ul{k'} \cdot (\ul{x} - \ul{y})} \,
  \bigg\{ f_1^N (x , |\ul{q} -\ul{k}' - x \, \ul{p}|_T) \notag \\
  & \times \, A \, \bigg[ W_{unp}^{OAM} \left(p^+, \ul{p} , b^-,
    \frac{\ul{x} + \ul{y}}{2} \right) \, S_{{\ul{y} \, \ul{x}}} [b^-,
  -\infty] + W_{unp}^{symm} \left(p^+, - \ul{p} , b^-, \frac{\ul{x} +
      \ul{y}}{2} \right) \, i \, O_{{\ul{y} \, \ul{x}}} [b^-, -\infty]
  \bigg] \notag \\ & + \frac{1}{m_N} {\hat z} \cdot \left(\ul{S}
    \times (\ul{q} -\ul{k}'-x \, \ul{p})\right) \, f_{1T}^{\perp N} (x
  , |\ul{q} -\ul{k}'- x \ul{p}|_T) \\
  & \times \bigg[ W_{trans}^{symm} \left(p^+, \ul{p}, b^-,
    \frac{\ul{x} + \ul{y}}{2} \right) \, S_{{\ul{y} \, \ul{x}}} [b^-,
  -\infty] + W_{trans}^{OAM} \left(p^+, - \ul{p}, b^-, \frac{\ul{x} +
      \ul{y}}{2} \right) \, i \, O_{{\ul{y} \, \ul{x}}} [b^-, -\infty]
  \bigg] \bigg\}. \notag
\end{align}
As before, we drop contributions from the odderon $iO_{{\ul{y} \,
    \ul{x}}}$ as being outside the precision of the quasi-classical
formula \eqref{DYTMD1} to get
\begin{align} 
  \label{DY_TMD7} & {\hat z} \cdot (\ul{J}\times\ul{q}) \,
  f_{1T}^{\perp A} ( {\bar x} , q_T) = \, M_A \int \frac{d p^+ \, d^2
    p \, d b^-}{2(2\pi)^3} \int \frac{d^2 k' \, d^2 x \,
    d^2 y}{(2\pi)^2} \, e^{i \ul{k'} \cdot (\ul{x} - \ul{y})} \notag \\
  & \times \, \bigg\{ A \, W_{unp}^{OAM} \left(p^+, \ul{p} , b^-,
    \frac{\ul{x} + \ul{y}}{2} \right) \, f_1^N (x , |\ul{q} -\ul{k}' -
  x \, \ul{p}|_T) \\ & + \frac{1}{m_N} {\hat z} \cdot \left(\ul{S}
    \times (\ul{q} -\ul{k}'-x \, \ul{p})\right) \, W_{trans}^{symm}
  \left(p^+, \ul{p}, b^-, \frac{\ul{x} + \ul{y}}{2} \right) \,
  f_{1T}^{\perp N} (x , |\ul{q} -\ul{k}'- x \ul{p}|_T) \bigg\} \,
  S_{{\ul{y} \, \ul{x}}} [b^-, -\infty]. \notag
\end{align}

Since the rotational momentum of the nucleons $p_T$ is assumed to be
small, we have to expand in it to the lowest non-trivial
order. Shifting the integration variable $\ul{k}' \to \ul{k}' + \ul{q}
- x \, \ul{p}$ in \eq{DY_TMD7} and expanding the exponential to the
lowest non-trivial order in $p_T$ we obtain (cf. \eq{TMD15})
\begin{align} 
  \label{DY_TMD8} & {\hat z} \cdot (\ul{J}\times\ul{q}) \,
  f_{1T}^{\perp A} ( {\bar x} , q_T) = \, M_A \int \frac{d p^+ \, d^2
    p \, d b^-}{2(2\pi)^3} \int \frac{d^2 k' \, d^2 x \,
    d^2 y}{(2\pi)^2} \, e^{- i \, (\ul{q} - \ul{k'}) \cdot (\ul{x} - \ul{y})} \notag \\
  & \times \, \bigg\{ i \, x \, \ul{p} \cdot (\ul{x} - \ul{y}) \, A \,
  W_{unp}^{OAM} \left(p^+, \ul{p} , b^-, \frac{\ul{x} + \ul{y}}{2}
  \right) \, f_1^N (x , k'_T) \\ & - \frac{1}{m_N} {\hat z} \cdot
  \left(\ul{S} \times \ul{k}' \right) \, W_{trans}^{symm} \left(p^+,
    \ul{p}, b^-, \frac{\ul{x} + \ul{y}}{2} \right) \, f_{1T}^{\perp N}
  (x , k'_T) \bigg\} \, S_{{\ul{x} \, \ul{y}}} [b^-, -\infty], \notag
\end{align}
where we have also interchanged $\ul{x} \leftrightarrow \ul{y}$ and
$\ul{k}' \to - \ul{k}'$.

\eq{DY_TMD8} is our main formal result for the DY Sivers function. We
again see that the Sivers function in DY can arise through two
distinct channels in this quasi-classical approach: the OAM channel
that contains its preferred direction in the distribution
$W_{unp}^{OAM}$ and the transversity/Sivers density channel that
generates is preferred direction through a local lensing mechanism
$f_{1T}^{\perp N}$.

To demonstrate the importance of the Wilson lines for the Sivers
function, for the moment, let us ignore the contribution of the Wilson
lines associated with initial-state rescattering in
\eq{DY_TMD8}. Without any such initial-state interactions, the
nucleonic Sivers function is zero, $f_{1T}^N = 0$
\cite{Brodsky:2002rv,Boer:2002ju,Brodsky:2013oya}, leaving
\begin{align}
  \label{DY10}
  {\hat z} \cdot (\ul{J}\times\ul{q}) \, f_{1T}^{\perp A} ( {\bar x} ,
  q_T) = \, M_A \int \frac{d p^+ \, d^2 p \, d b^-}{2(2\pi)^3} \int
  \frac{d^2 k' \, d^2 x \,
    d^2 y}{(2\pi)^2} \, e^{- i \, (\ul{q} - \ul{k'}) \cdot (\ul{x} - \ul{y})} \notag \\
  \times \, i \, x \, \ul{p} \cdot (\ul{x} - \ul{y}) \, A \,
  W_{unp}^{OAM} \left(p^+, \ul{p} , b^-, \frac{\ul{x} + \ul{y}}{2}
  \right) \, f_1^N (x , k'_T) = 0,
\end{align}
which vanishes after $b^-$ integration because of the rotational and
$PT$-symmetry conditions \eqref{rot6}.


\subsection{Sign Reversal of the Sivers Function Between SIDIS and DY}

Now that the DY Sivers function \eqref{DY_TMD8} is expressed in the
same form as the Sivers function for SIDIS \eqref{TMD15}, we can
compare both expression to see how the nuclear Sivers functions have
changed between SIDIS and DY and understand the origin of the SIDIS/DY
sign-flip relation \cite{Collins:2002kn}
\begin{align} 
  \label{DYTMD9} f_{1T}^{\perp A} (x , k_T)\bigg|_{SIDIS} = -
  f_{1T}^{\perp A} (x , k_T)\bigg|_{DY}.
\end{align}
First, we notice that the transversity/Sivers density channel (the
second term in the curly brackets) has changed signs as required
between \eqref{TMD15} and \eqref{DY_TMD8}.  Mathematically, this
occurs because of the $\ul{k}' \to - \ul{k}'$ interchange, simply
because the momentum going into the Wilson line in SIDIS corresponds
to the momentum coming from the Wilson line in DY
(cf. Figs.~\ref{fig:TMD1} and \ref{fig:DY2}). The transversity/Sivers
density channel contribution thus automatically satisfies the
sign-flip relation \eqref{DYTMD9}.

The OAM channel contribution to \eq{DY_TMD8} is more subtle; although
the prefactor has not changed as compared to \eq{TMD15}, the
longitudinal coordinate $b^-$ integral entering \eqref{DY_TMD8} for DY
can be modified using $b^- \to - b^-$ substitution along with
\eq{rot6} to give
\begin{align}
  \label{DY_TMD11}
  \int db^- \, W_{unp}^{OAM} (p,b) \, S_{{\ul{x} \,
         \ul{y}}} [b^-, -\infty] = - \int db^- \, W_{unp}^{OAM} (p,b) \, S_{{\ul{x} \,
         \ul{y}}} [- b^-, -\infty]. 
\end{align}
When evaluating the dipole $S$-matrix we neglect the polarization
effects as being energy suppressed. Therefore, for the purpose of this
$S$-matrix, the nucleus has a rotational symmetry around the $z$-axis
(see \fig{nucleus} for axes labels). We thus write
\begin{align}
  \label{DY_TMD12}
  S_{{\ul{x} \, \ul{y}}} [- b^-, -\infty] \overset{PT}{=} S_{-{\ul{x},
      \, -\ul{y}}} [+ \infty, b^-] \overset{z-rotation}{=} S_{{\ul{x}
      \, \ul{y}}} [+ \infty, b^-],
\end{align}
where $z$-rotation denotes a half-revolution around the
$z$-axis. Using \eq{DY_TMD12} in \eq{DY_TMD11} we arrive at
\begin{align} \label{DYTMD10}
 \begin{aligned}
   \overbrace{\int db^- \, W_{unp}^{OAM}(p,b) \, S_{{\ul{x} \,
         \ul{y}}} [b^-, -\infty]}^{\mathrm{DY}} &= - \overbrace{\int
     db^- \, W_{unp}^{OAM} (p,b) \, S_{{\ul{x} \, \ul{y}}} [+ \infty,
     b^-] }^{\mathrm{SIDIS}}.
 \end{aligned}
\end{align}

One can also simply see that \eq{DYTMD10} is true by using the
quasi-classical GM/MV dipole $S$-matrix from \eq{Wlines4} on its
right-hand-side, along with
\begin{align} 
  \label{Wlines42} S_{{\ul x} {\ul y}}[b^-, -\infty] = \exp
  \left[-\frac{1}{4} |\ul{x}- \ul{y}|^2 \, Q_s^2 \left(\frac{{\ul x} +
        {\ul y}}{2} \right) \, \left(\frac{b^- + R^-}{2 R^-}\right) \,
    \ln\frac{1}{|\ul{x} -\ul{y}| \Lambda} \right]
\end{align}
on its left-hand-side. We conclude that the OAM channel contributions
to the SIDIS Sivers function \eqref{TMD15} and the DY Sivers function
\eqref{DY_TMD8} also satisfy the sign-flip relation \eqref{DYTMD9}.

Therefore, for any Wigner distribution $W(p,b)$, the Sivers functions
at the quasi-classical level for SIDIS \eqref{TMD15} and for DY
\eqref{DY_TMD8} are equal in magnitude and opposite in sign,
\eqref{DYTMD9}.  This statement is a direct consequence of the
invariance of $W(p,b)$ under rotations and $PT$-reversal,
\eqref{rot6}, and it mirrors in this context the original derivation
by Collins \cite{Collins:2002kn}.

The advantage of our approach here, apart from providing the explicit
formal results \eqref{TMD15} and \eqref{DY_TMD8}, is in the new
physical interpretation of the transverse spin asymmetry in the OAM
channel. As described in the Introduction following \fig{DY_fig}, the
incoming antiquark is more likely to interact with the ``front'' of
the nucleus, due to shadowing effects, thus scattering on the nucleon
moving out of the page in \fig{DY_fig}. This is justified by the
$S_{{\ul x} {\ul y}}[b^-, -\infty]$ function in \eq{DY_TMD8} (see also
\eqref{Wlines42}), which is largest for $b^- = - R^-$. Thus the
virtual photon is produced preferentially out of the page; this leads
to a non-zero STSA in DY. The sign reversal relation follows from
comparing Figs.~\ref{DY_fig} and \ref{SIDIS_fig}: in DY the particles
are produced preferentially left-of-beam, while in SIDIS the produced
hadrons come out mainly right-of-beam.

The rigid-rotator toy model of Sec.~\ref{toy_model} can also be
constructed for DY Sivers function. However, due to the sign-reversal
relation \eqref{DYTMD9} we can read off the answer for the DY Sivers
function in the rigid-rotator model as being negative of that in
\eq{SiversA4} for moderate $k_T$ and negative of \eq{eq:Sivers_highk}
for $k_T \gg Q_s$. All the conclusions about the relative importance
of the two contributing channels remain the same.


\section{Discussion}
\label{sec:Discussion}

The main goal of this work was to construct SIDIS and DY Sivers
functions in the quasi-classical GM/MV approximation, which models a
proton as a large nucleus, and which we modified by giving the nucleus
a non-zero OAM. The main formal results are given in
Eqs.~\eqref{TMD15} (SIDIS) and \eqref{DY_TMD8} (DY). We showed that
there are two main mechanisms generating the quasi-classical Sivers
function: the OAM channel and the transversity channel. The former is
leading in saturation power counting; it also dominates for $k_T <
Q_s/\sqrt{\as}$, that is both inside and, for $Q_s < k_T <
Q_s/\sqrt{\as}$, outside of the saturation region. At higher $k_T$ the
transversity channel dominates. In the future our quasi-classical
calculation can be augmented by including evolution corrections to the
Sivers function, making the whole formalism ready for phenomenological
applications, similar to the successful use of nonlinear small-$x$
evolution equations
\cite{Balitsky:1996ub,Kovchegov:1999yj,Jalilian-Marian:1997dw,Iancu:2000hn,Balitsky:2006wa,Gardi:2006rp,Kovchegov:2006vj}
to description (and prediction) of high energy scattering data
\cite{Albacete:2010sy,ALbacete:2010ad}.

Perhaps just as important, we constructed a novel physical mechanism
of the STSA generation. This is the OAM channel. The OAM mechanism,
while diagrammatically very similar to the original BHS mechanism
\cite{Brodsky:2002cx}, provides a different interpretation from the
'lensing' effect \cite{Brodsky:2002cx,Brodsky:2013oya} or the
color-Lorentz force of \cite{Burkardt:2010sy,Burkardt:2008ps}. The OAM
mechanism is described in the Introduction, in the discussion around
Figs.~\ref{DY_fig} and \ref{SIDIS_fig}. It is based on interpreting
the extra rescattering proposed by BHS as a shadowing-type
correction. The STSA is then generated by the combination of the OAM
and shadowing. The shadowing makes sure the projectile interacts
differently with the front and the back of the target, generating the
asymmetry of the produced particles.

While shadowing is a high-energy phenomenon, and our calculation was
done in the high-energy approximation ${\hat s} \gg \perp^2$ (though
for $x \sim \ord{1}$), it may be that the OAM mechanism for generating
STSA is still valid for lower-energy scattering, though of course the
formulas derived above would not apply in such regime. At lower
energies the difference between the interactions of the projectile
with the front/back of the target may result from, say, energy loss of
the projectile as it traverses the target. Again, combined with the
target rotation this would generate STSA, and, hence, the Sivers
function. The formalism needed to describe such a low-energy process
would be quite different from the one presented above; moreover, the
correct degrees of freedom may not be quarks and gluons
anymore. However, the main physics principle of combining OAM with the
difference in interaction probabilities between the projectile and
front/back of the target to generate STSA may be valid at all
energies.

Returning to higher energies and the derived formulas~\eqref{TMD15}
and \eqref{DY_TMD8}, let us point out that these results, when applied
to experimental data, may allow one to determine the distribution of
intrinsic transverse momentum $\ul{p} (\ul{b}, b^-)$ of partons in the
hadronic or nuclear target, along with the transversity/Sivers
function density in the target. This would complement the existing
methods of spatial imaging of quarks and gluons inside the hadrons and
nuclei \cite{Accardi:2012qut}, providing a new independent cross-check
for those methods.


\section*{Acknowledgments}

We would like to thank Daniel Boer, Dick Furnstahl, Leonard Gamberg
and Feng Yuan for useful correspondence.

This research is sponsored in part by the U.S. Department of
Energy under Grant No. DE-SC0004286.


\appendix

\renewcommand{\theequation}{A\arabic{equation}}
\setcounter{equation}{0}
\section{Wigner Distributions with Multiple Rescatterings}
\label{A}

\begin{figure}[ht]
\centering
\includegraphics[height=4cm]{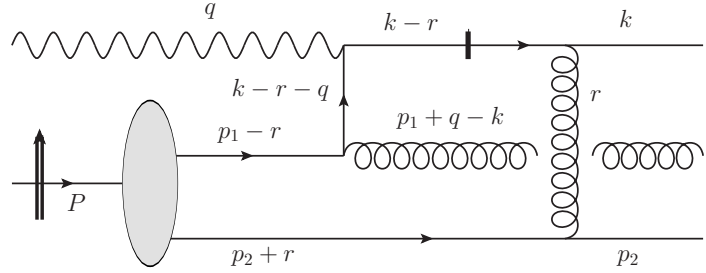}
\caption{
  \label{fig:DIS2} The minimal SIDIS process containing both the
  ``knockout'' of a quark from the nuclear wave function and
  rescattering on a different quark from a second nucleon.  The short
  thick vertical line indicates that the pole of the intermediate
  quark propagator is picked up in the calculation.  }
\end{figure}

The aim of this Appendix is to justify the result given in
\eq{eq:ampl2_LO_Wxy_net}. To study the interplay between the local
``knockout'' channel of deep inelastic scattering and the coherent
multiple rescattering on the nuclear remnants, it is illustrative to
consider a minimal case with both features.  This process, shown in
Fig.~\ref{fig:DIS2}, consists of the knockout sub-process followed by
a single rescattering on a different quark from a second nucleon in
the nucleus.  Rescattering on a second nucleon receives a combinatoric
enhancement of order $\sim A^{1/3}$ compared to rescattering on the
same nucleon; the former is $\ord{1}$ in the saturation power
counting, while the latter is $\ord{\alpha_s}$.

The total SIDIS amplitude $M_{tot}$ depicted in Fig.~\ref{fig:DIS2}
consists of a loop integral connecting the mean-field single-particle
wave functions $\psi(p)$ of the nucleus to a scattering amplitude
$M_{K+R}$ denoting both the ``knockout'' and rescattering processes:
\begin{align} 
\label{DIS6} 
M_{tot} = \int \frac{d r^+ \, d^2r }{2 \, (2 \pi)^3} \frac{P^+}{(p_1^+
  - r^+) \, (p_2^+ + r^+)} \psi(p_1 -r) \, \psi(p_2 + r) \,
M_{K+R}(p_1 - r,p_2 + r,q,k, r),
\end{align}
where a sum over spins and colors of the participating quarks is
implied. Squaring both sides of \eqref{DIS6} and integrating out the
final-state momenta $p_1$ and $p_2$ gives
\begin{align} \label{DIS7}
  \langle |M_{tot}|^2 \rangle & \equiv A \, (A -1) \int \frac{d p_1^+
    \, d^2 p_1 \, dp_2^+ \, d^2p_2}{[2 \, (2 \pi)^3]^2 \, (p_1^+ +
    q^+) \, p_2^+ } |M_{tot}|^2 \\ = & \int \frac{d p_1^+ \, d^2 p_1 \,
    dp_2^+ \, d^2p_2}{[2 \, (2 \pi)^3]^2 \, (p_1^+ + q^+) \, p_2^+ }
  \frac{d r^+ \, d^2 r}{2 \, (2 \pi)^3} \frac{d r^{\prime +} \, d^2
    r'}{2 \, (2 \pi)^3} \frac{A \, (A -1) \, \left( P^+
    \right)^2}{\sqrt{(p_1^+ - r^+) \, (p_2^+ + r^+) \,
      (p_1^+ - r^{\prime +})  \, (p_2^+ + r^{\prime +})}} \notag \\
  & \times \int d b_1^- \, d^2 b_1 \, d b_2^- \, d^2 b_2 \, e^{- i \,
    (r - r')\cdot(b_1 - b_2)} \, W \left(p_1 - \frac{r + r'}{2},b_1
  \right)
  W \left(p_2 + \frac{r + r'}{2} , b_2 \right) \notag \\
  & \times M_{K+R}(p_1 - r, p_2 + r,q,k, r) \, M^*_{K+R}(p_1 - r',p_2
  + r',q,k, r'), \notag
\end{align}
where we have employed the Wigner distributions defined in \eq{DIS8}
above and summed over all pairs of nucleons.

\eq{DIS7} is still far from \eq{eq:ampl2_LO_Wxy_net} because in
\eqref{DIS7} we do not have the amplitude squared: instead we have the
product of $M_{K+R}$ and $M^*_{K+R}$ with different arguments. It is
easier to further analyze the expression separately for the transverse
and longitudinal degrees of freedom. We proceed by taking the
classical limits, in which case the Wigner distributions give us the
position and momentum distributions of nucleons
simultaneously. Moreover, for the large nucleus at hand the Wigner
distributions depend on ${\ul b}_1$ and ${\ul b}_2$ weakly over the
perturbatively short distances associated with the Feynman
diagrams. We thus define ${\ul b} = ({\ul b}_1 + {\ul b}_2)/2$ and
${\ul \Delta b} = {\ul b}_1 - {\ul b}_2$ and write
\begin{align}
  \label{eq:trans}
  & \int d^2 r \, d^2 r' \, d^2 b_1 \, d^2 b_2 \, e^{i \, ({\ul r} -
    {\ul r}')\cdot({\ul b}_1 - {\ul b}_2)} \, W \left(p_1 - \frac{r +
      r'}{2},b_1 \right) W \left(p_2 + \frac{r + r'}{2} , b_2 \right)
  \notag \\ & \times M_{K+R}(p_1 - r, p_2 + r,q,k, r) \, M^*_{K+R}(p_1
  - r',p_2 + r',q,k, r') \approx \int d^2 r \, d^2 r' \, d^2 b \, d^2
  \Delta b\, e^{i \, ({\ul r} - {\ul r}')\cdot {\ul \Delta b}} \notag
  \\ & \times W \left(p_1 - \frac{r + r'}{2}, b_1^-, {\ul b} \right) W
  \left(p_2 + \frac{r + r'}{2} , b_2^-, {\ul b}\right) \, M_{K+R}(p_1
  - r, p_2 + r,q,k,r ) \notag \\ & \times M^*_{K+R}(p_1 - r',p_2 +
  r',q,k, r') = (2 \pi)^2 \, \int d^2 r \, d^2 b \, W \left(p_1^+ -
    \frac{r^+ + r^{\prime +}}{2}, {\ul p}_1 - {\ul r}, b_1^-, {\ul b}
  \right) \notag \\ & \times W \left(p_2^+ + \frac{r^+ + r^{\prime
        +}}{2}, {\ul p}_2 + {\ul r} , b_2^-, {\ul b} \right) \,
  M_{K+R}(p_1^+ - r^+, {\ul p}_1 - {\ul r}, p_2^+ + r^+, {\ul p}_2 +
  {\ul r} ,q,k, r^+, {\ul r}) \notag \\ & \times M^*_{K+R}(p_1^+ -
  r^{\prime +}, {\ul p}_1 - {\ul r}, p_2^+ + r^{\prime +}, {\ul p}_2 +
  {\ul r} , q,k, r^{\prime +}, {\ul r}).
\end{align}

Now the difference in the arguments of $M_{K+R}$ and $M^*_{K+R}$ is
only in the longitudinal momenta $r^+$ and $r^{\prime +}$. To
integrate over these momenta we notice that, as follows from
\fig{fig:DIS2}, in the high energy kinematics at hand the leading
contribution to the amplitude $M_{K+R}$ comes from the region where
$p_1^+, p_2^+ \gg r^+, r^{\prime +}$. In this regime we combine
Eqs.~\eqref{DIS7} and \eqref{eq:trans} to write
\begin{align}
  \label{eq:app1}
  \langle |M_{tot}|^2 \rangle & = \int \frac{d p_1^+ \, d^2 p_1 \,
    dp_2^+ \, d^2p_2}{[2 \, (2 \pi)^3]^2 \, (p_1^+ + q^+) \, p_2^+ }
  \frac{d r^+ \, d r^{\prime +} \, d^2 r}{4 \, (2 \pi)^4} \frac{\left(
      P^+ \right)^2}{p_1^+ \, p_2^+} d b_1^- \, d b_2^- d^2 b \, e^{-
    i \tfrac{1}{2} \, ( r^+ - r^{\prime +}) \, (b_1^- - b_2^-)} A \,
  (A -1) \notag \\ & \times \, W \left(p_1^+, {\ul p}_1 - {\ul r},
    b_1^-, {\ul b} \right) W \left(p_2^+, {\ul p}_2 + {\ul r} , b_2^-,
    {\ul b} \right) \, M_{K+R}(p_1^+, {\ul p}_1 - {\ul r}, p_2^+, {\ul
    p}_2 + {\ul r} ,q,k, r^+, {\ul r}) \notag \\ & \times \,
  M^*_{K+R}(p_1^+, {\ul p}_1 - {\ul r}, p_2^+, {\ul p}_2 + {\ul r} ,
  q,k, r^{\prime +}, {\ul r}).
\end{align}
In the $p_1^+, p_2^+ \gg r^+, r^{\prime +}$ kinematics the amplitude
$M_{K+R}$ contains only one pole in $r^+$ resulting from the
denominator of the $k-r$ quark propagator
(cf. \cite{Mueller:1989st,Kovchegov:1998bi,KovchegovLevin}). We can
thus write
\begin{align}
  \label{eq:ampl}
  M_{K+R}(p_1 - r, p_2 + r,q,k) = \frac{i}{(k-r)^2 + i \, \epsilon} \,
  {\tilde M}_{K+R}(p_1 - r, p_2 + r,q,k),
\end{align}
where ${\tilde M}_{K+R}$ denotes the rest of the diagram which does
not contain singularities in $r^+$ in the $p_1^+, p_2^+ \gg r^+,
r^{\prime +}$ approximation. (Note that ${\tilde M}_{K+R}$ also
contains the numerator of the $k-r$ quark propagator.) Since $(k-r)^2
\approx - k^- \, r^+ + {\ul k}^2 - ({\ul k} - {\ul r})^2$ we can use
\eq{eq:ampl} to integrate over $r^+$,
\begin{align}
  \label{eq:long_int}
  \int\limits_{-\infty}^\infty \frac{d r^+}{2 \pi} \, & e^{- i
    \tfrac{1}{2} \, r^+ \, (b_1^- - b_2^-)} \, M_{K+R}(p_1 - r, p_2 +
  r,q,k) \notag \\ & \approx \frac{1}{k^-} \, \theta (b_2^- - b_1^-)
  \, {\tilde M}_{K+R} (p_1^+ , {\ul p}_1 - {\ul r}, p_2^+ , {\ul p}_2
  + {\ul r} ,q,k) \notag \\ & = \frac{1}{k^-} \, \theta (b_2^- -
  b_1^-) \, M_K(p_1 -r, q, k-r) \, M_R(p_2+r, k-r, k, r).
\end{align}
Here we assumed that $r^+ = \left[ {\ul k}^2 - ({\ul k} - {\ul r})^2
\right]/k^- \approx 0$ in our kinematics. After putting the $k-r$
quark propagator on mass shell the amplitude ${\tilde M}_{K+R}$
factorizes into a product of separate amplitudes for knockout $M_K(p_1
-r, q, k-r)$ and rescattering $M_R(p_2+r, k-r, k, r)$
\cite{Mueller:1989st,Kovchegov:1998bi,KovchegovLevin}, as employed in
\eq{eq:long_int}, where the sum over quark polarizations and colors is
implicit.

With the help of \eq{eq:long_int} (and a similar one for the
$r^{\prime +}$-integration of $M^*_{K+R}$) we write
\begin{align}
  \label{eq:app2}
  \langle |M_{tot}|^2 \rangle = \int \frac{d p_1^+ \, d^2 p_1 \,
    dp_2^+ \, d^2p_2}{[2 \, (2 \pi)^3]^2 \, (p_1^+ + q^+) \, p_2^+ }
  \frac{d^2 r}{4 \, (2 \pi)^2} \frac{A \, (A-1) \, \left( P^+
    \right)^2}{p_1^+ \, p_2^+ \, (k^-)^2} \, d b_1^- \, d b_2^- \, d^2
  b \, \theta (b_1^- - b_2^-) \notag \\ \times \, W \left(p_1^+, {\ul
      p}_1 - {\ul r}, b_1^-, {\ul b} \right) \, W \left(p_2^+, {\ul
      p}_2 + {\ul r} ,
    b_2^-, {\ul b} \right) \notag \\
  \times \, |M_K(p_1 -r, q, k-r)|^2 \, |M_R(p_2+r, k-r, k, r)|^2.
\end{align}

Defining the energy-independent (at the quasi-classical level)
rescattering amplitude by \cite{Mueller:1989st,KovchegovLevin}
\begin{align}
  \label{eq:e-ind_ampl}
  | A_R (p_2+r, k-r, k, r)|^2 \equiv \frac{1}{4 (p_2^+)^2 \, (k^-)^2}
  \, |M_R(p_2+r, k-r, k, r)|^2
\end{align}
and denoting the average of this object in the Wigner distribution by
the angle brackets
\begin{align}
  \label{eq:Wave}
  \left\langle | A_R (k, r)|^2 \right\rangle (b_1^-, {\ul b}) = \int
  \frac{dp_2^+ \, d^2p_2 \, d b_2^-}{2 \, (2 \pi)^3} \, \theta (b_2^-
  - b_1^-) \, (A-1) \, W \left(p_2^+, {\ul p}_2 + {\ul r} , b_2^-,
    {\ul b} \right) \notag \\ \times \, |A_R(p_2+r, k-r, k, r)|^2
\end{align}
we rewrite \eq{eq:app2} as
\begin{align}
  \label{eq:app3}
  \langle |M_{tot}|^2 \rangle = & A \, \int \frac{d p_1^+ \, d^2 p_1
    \, d b_1^- \, d^2 b}{2 \, (2 \pi)^3} \, \frac{\left( P^+
    \right)^2}{p_1^+ \, (p_1^+ + q^+)} \, W \left(p_1^+, {\ul p}_1,
    b_1^-, {\ul b} \right) \notag \\ & \times \, \int \frac{d^2 r}{(2
    \pi)^2} \, |M_K(p_1, q, k -r)|^2 \, \left\langle | A_R (k, r)|^2
  \right\rangle (b_1^-, {\ul b}).
\end{align}
In arriving at \eq{eq:app3} we have shifted the momentum $p_1 \to p_1
+ r$. 

We now define the ``energy-independent'' total and ``knockout''
amplitudes \cite{Mueller:1989st,KovchegovLevin}
\begin{align}
  \label{eq:A_Eindep}
  |A_{tot}|^2 \equiv \frac{1}{4 \, \left( P^+ \right)^2 \, (q^-)^2} \,
  |M_{tot}|^2, \ \ \ \ \ |A_k|^2 \equiv \frac{1}{4 \, ( p_1^+ )^2 \,
    (q^-)^2} \, |M_{K}|^2.
\end{align}
Employing the Fourier transform \eqref{M_Ftr} we reduce \eq{eq:app3}
to
\begin{align}
  \label{eq:app4}
  \langle |A_{tot}|^2 \rangle = & A \int \frac{d p_1^+ \, d^2 p_1 \, d
    b_1^- \, d^2 b}{2 \, (2 \pi)^3} \, \frac{p_1^+}{p_1^+ + q^+} \, W
  \left(p_1^+, {\ul p}_1, b_1^-, {\ul b} \right) \, \int d^2 x \, d^2
  y \, e^{- i \, {\ul k} \cdot ({\ul x} - {\ul y})} \notag \\ & \times
  \, A_K(p_1, q, k^-, r^+, {\ul x} - {\ul b}) \, A_K^* (p_1, q, k^-,
  r^+, {\ul y} - {\ul b}) \, \left\langle | A_R|^2 \right\rangle (k^-,
  {\ul x} - {\ul y}, b_1^-, {\ul b})
\end{align}
with
\begin{align}
  \label{eq:AR_FT}
  \left\langle | A_R|^2 \right\rangle (k^-, {\ul x} - {\ul y}, b_1^-,
  {\ul b}) = \int \frac{d^2 r}{(2 \pi)^2} \, e^{i \, {\ul r} \cdot ({\ul
      x} - {\ul y})} \,\left\langle | A_R (k, r)|^2 \right\rangle
  (b_1^-, {\ul b}).
\end{align}

Comparing \eq{eq:app3} to \eq{eq:ampl2_LO_Wxy} we see that, just like
in all high energy QCD scattering calculations
\cite{Mueller:1989st,KovchegovLevin,Weigert:2005us,Jalilian-Marian:2005jf,Gelis:2010nm}
the rescattering can be factored out into a multiplicative factor in
the transverse coordinate space. Similar to the above one can show that
all further rescatterings would only introduce more multiplicative
factors. Defining a somewhat abbreviated notation
\begin{align}
  \label{eq:M2_def}
  A (p, q, {\ul x} - {\ul b}) \, A^* (p, q, {\ul y} - {\ul b}) \equiv
  A_K(p, q, k^-, r^+, {\ul x} - {\ul b}) \, A_K^* (p, q, k^-, r^+,
  {\ul y} - {\ul b}) \notag \\ \times \, \left\langle | A_R|^2
  \right\rangle (k^-, {\ul x} - {\ul y}, b_1^-, {\ul b})
\end{align}
we see that \eq{eq:app4} reduces to \eq{eq:ampl2_LO_Wxy_net}, as
desired. The above discussion also demonstrates how multiple
rescatterings factorize in the transverse coordinate space: in the
high energy kinematics they are included through the Wilson lines of
Eqs.~\eqref{xsectNNN} and \eqref{dipole_def}. The Wilson line
correlator $D_{{\ul x} \, {\ul y}} [+\infty, b^-]$ from
\eqref{dipole_def} contains a $b^-$-ordered product of multiple
rescattering factors $\left\langle | A_R|^2 \right\rangle$ from all
the interacting nucleons
\cite{Jalilian-Marian:1997xn,Kovchegov:1998bi}.


\renewcommand{\theequation}{B\arabic{equation}}
 \setcounter{equation}{0}
 \section{The Role of $PT$-Symmetry }
\label{B}

The decompositions \eqref{TMD15} and \eqref{DY_TMD8} essentially break
the Wilson line operator $\mathcal{U}$ in the definition
\eqref{eq:q_corr} into two parts: the coherent rescattering $S_{{\ul
    x} {\ul y}}[+\infty, b^-]$ on other spectator nucleons which is a
leading-order contribution in the saturation power counting, and the
subleading lensing interaction with the same nucleon which generates
$f_{1T}^{\perp N}$.  If we neglect the Wilson line operator
$\mathcal{U}$ entirely, then we know that the Sivers function of the
nucleus $f_{1T}^{\perp A}$ must vanish, as first proved by Collins
\cite{Collins:1992kk}. But if we drop $f_{1T}^{\perp N}$ and $S_{{\ul
    x} {\ul y}}[+\infty, b^-]$ from \eqref{TMD15}, we do not obviously
get zero:
\begin{figure}[h]
\centering
\includegraphics[height=6cm]{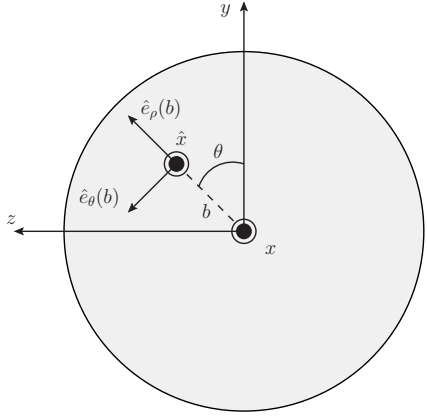}
\caption{
  \label{fig:W_Rot} Definition of the cylindrical coordinate basis
  \eqref{rot1} convenient for formulating the symmetry properties of
  the nucleonic distribution $W_\sigma(p,b)$ in the rest frame of the
  nucleus.  }
\end{figure}
%
\begin{align} \label{TMD16}
  \begin{aligned} {\hat z} \cdot & (\ul{J}\times\ul{k}) \,
    f_{1T}^{\perp A} (x , k_T) = M_A \, A \int \frac{d p^+ \, d^2 p \,
      d b^-}{2(2\pi)^3} \, d^2x \, d^2 y \, \frac{d^2 k'}{(2\pi)^2} \,
    e^{ - i \, (\ul{k} - \ul{k'})
      \cdot(\ul{x}-\ul{y})}   \\
    & \times \, i \, x \, \ul{p} \cdot (\ul{x}-\ul{y}) \,
    W_{unp}^{OAM} \left( p^+, \ul{p} ,b^-, \frac{{\ul x} + {\ul y}}{2}
    \right) \, f_1^N (x, k'_T) \stackrel{?}{=} 0.
 \end{aligned}
\end{align}
The right-hand side of this equation must vanish for wave functions
described by $W_{unp}^{OAM}$ that are $PT$ eigenstates
\cite{Collins:1992kk}; we can see this explicitly by considering the
constraints on $W_\sigma(p,b)$ due to rotational invariance and $PT$
symmetry.  It is most convenient to enumerate the rotational symmetry
properties of the nucleon distribution $W_\sigma(p,b)$ in the rest
frame of the nucleus, using a cylindrical vector basis coaxial to the
transverse spin vector $\ul{S}$.  This basis $(\hat{e}_\rho ,
\hat{e}_\theta , \hat{x})$ is shown in Fig.~\ref{fig:W_Rot} and is
defined by
\begin{align} \label{rot1}
 \begin{pmatrix} \hat{e}_\rho \\ \hat{e}_\theta \end{pmatrix} =
 \begin{pmatrix}
  b_y / b_\rho & b_z / b_\rho \\
  -b_z/ b_\rho & b_y / b_\rho 
 \end{pmatrix}
 \begin{pmatrix} \hat{y} \\ \hat{z} \end{pmatrix} =
 \begin{pmatrix}
  \cos\theta & \sin\theta \\
  -\sin\theta & \cos\theta 
 \end{pmatrix}
 \begin{pmatrix} \hat{y} \\ \hat{z} \end{pmatrix}
\end{align}
where $\left( p_\rho(b) , p_\theta(b) \right) = p \cdot
\left(\hat{e}_\rho(b) , \hat{e}_\theta(b)\right)$ and $b_\rho \equiv
\sqrt{b_y^2 + b_z^2}$.

\begin{figure}
\centering
\includegraphics[width=\textwidth]{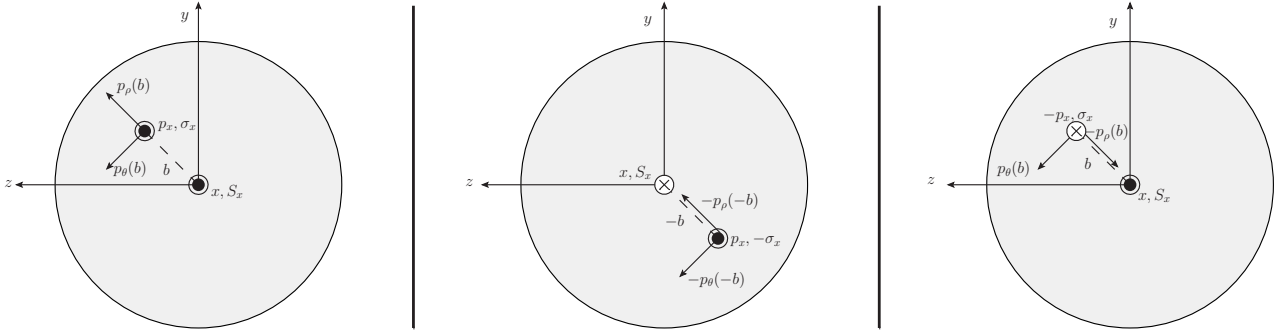}
\caption{
  \label{fig:PT_Rot} Illustration of the $PT$ transformation and
  rotational symmetry in the rest frame used in \eqref{rot3}.  Left
  panel: illustration of the momentum flow represented by
  $W_\sigma(p,b)$.  Center panel: under a $PT$ transformation, the
  spins $S,\sigma$ and coordinate $b$ are reversed, but the momentum
  $p$ is invariant.  Right panel: rotation of the center panel by
  $\pi$ about the vector $\vec{S}\times\vec{b}$ returns the
  distribution to its original position $b$, with $p_\rho$ and $p_x$
  having been reversed.}
\end{figure}

First, the distribution must be symmetric under rotations about the
transverse spin $S_x$, which are easy to express in this cylindrical
basis:
\begin{align} \label{rot2}
 W_{\sigma} \big(p_x \, ; \, p_\rho (b) \, ; \, p_\theta (b) \, ; \, b \big) = 
 W_{\sigma} \big( p_x \, ; \, p_\rho (b') \, ; \, p_\theta (b') \, ; \, b' \big).
\end{align}
Second, if the nucleus is in a $PT$-symmetric eigenstate of the QCD
Hamiltonian, then $W_\sigma(p,b)$ should be invariant under $PT$
transformations.  These transformations reverse the coordinates $(b
\rightarrow -b)$ and pseudovectors like the spin $(S , \sigma
\rightarrow -S , -\sigma)$, but leave the momentum vector $p$
unchanged.  Using this transformation, together with rotational
invariance as shown in Fig.~\ref{fig:PT_Rot} we obtain
\begin{align} \label{rot3}
 \begin{aligned}
 W_\sigma \left( p_\rho(b) , p_\theta(b) , p_x ; b ; S_x \right) &\overset{PT}{=}
 W_{-\sigma} \left( p_\rho(b) , p_\theta(b) , p_x ; -b ; -S_x \right) \\ &=
 W_{-\sigma} \left( -p_\rho(-b) , -p_\theta(-b) , p_x ; b ; -S_x \right) \\ &\overset{R_b}{=}
 W_\sigma \left( -p_\rho(b) , p_\theta(b) , -p_x ; b ; S_x \right) \\ &\therefore \\
 W_\sigma \left( p_\rho(b) , p_\theta(b) , p_x ; b ; S_x \right) &= 
 W_\sigma \left( -p_\rho(b) , p_\theta(b) , -p_x ; b ; S_x \right),
 \end{aligned}
\end{align}
where the rotation $R_b$ is a half-revolution in the $Sb$-plane.  
%
%
%
This means that in a $PT$ eigenstate with transverse spin $S_x$, the
only allowed direction of net momentum flow corresponds to the
azimuthal orbital momentum $p_\theta$ and explains the naming
convention $W^{OAM}$ in \eqref{DIS21}.

The distributions that enter \eqref{TMD15} and \eqref{DY_TMD8},
however, are the (anti)symmetrized distributions under reversal of the
transverse momenta $(p_x, p_y \rightarrow -p_x , -p_y)$.  For these
purposes, it is more convenient to write the distribution $W_\sigma
(p,b)$ in terms of the Cartesian basis
\begin{align} \label{rot4} W(p_x , p_y, p_z ; b) = W_\sigma \left( p_x
    \,;\, \frac{b_y}{b_\rho} p_\rho(b) - \frac{b_z}{b_\rho}
    p_\theta(b) \,;\, \frac{b_z}{b_\rho} p_\rho(b) +
    \frac{b_y}{b_\rho} p_\theta(b) \,;\, b \right).
\end{align}
Using the symmetry properties \eqref{rot2} and \eqref{rot3}, we can
write the $\ul{p}$-reversed distribution in terms of the distribution
at a point $\overline{b} \equiv (b_x, b_y, -b_z)$ on the opposite side
of the nucleus:
\begin{align} \label{rot5}
 \begin{aligned}
   W_\sigma(-p_x , -p_y, &p_z ; b) = W_\sigma \left( -p_x \,;\,
     -\frac{b_y}{b_\rho} p_\rho(b) + \frac{b_z}{b_\rho} p_\theta(b)
     \,;\, \frac{b_z}{b_\rho} p_\rho(b) + \frac{b_y}{b_\rho}
     p_\theta(b) \,;\, b \right) \\ &\overset{Eq.\eqref{rot2}}{=}
   W_\sigma \left( -p_x \,;\, -\frac{b_y}{b_\rho} p_\rho(\overline{b})
     - \frac{b_z}{b_\rho} p_\theta(\overline{b}) \,;\, -
     \frac{b_z}{b_\rho} p_\rho(\overline{b}) + \frac{b_y}{b_\rho}
     p_\theta(\overline{b}) \,;\, \overline{b} \right) \\
   &\overset{Eq.\eqref{rot3}}{=} W_\sigma \left( p_x \,;\,
     \frac{b_y}{b_\rho} p_\rho(\overline{b}) - \frac{b_z}{b_\rho}
     p_\theta(\overline{b}) \,;\, \frac{b_z}{b_\rho}
     p_\rho(\overline{b}) + \frac{b_y}{b_\rho} p_\theta(\overline{b})
     \,;\, \overline{b} \right) \\ &=
   W_\sigma (p_x , p_y , p_z ; \overline{b}) \\ &\therefore \\
   W_\sigma(-p_x , -p_y, &p_z ; b) = W_\sigma (p_x , p_y , p_z ;
   \overline{b}).
\end{aligned}
\end{align}
Thus a nucleon on the back side of the nucleus has an opposite
transverse momentum to a corresponding nucleon in the front of the
nucleus.  Therefore, the (anti)symmetrized distributions have definite
parity under $b_z \rightarrow - b_z$:
\begin{align} \label{rot6}
 \begin{aligned}
   W_\sigma^{symm}(p,b) &\equiv \frac{1}{2} \left[ W_\sigma(p,b) +
     (\ul{p} \rightarrow - \ul{p}) \right] =
   + W_\sigma^{symm}(p,\overline{b}) \\
   W_\sigma^{OAM}(p,b) &\equiv \frac{1}{2} \left[ W_\sigma(p,b) -
     (\ul{p} \rightarrow - \ul{p}) \right] = -
   W_\sigma^{OAM}(p,\overline{b}).
 \end{aligned}
\end{align}

Eq.~\eqref{rot6} tells us that a consequence of $PT$ invariance in the
nucleus is that the orbital angular momentum encountered at any point
in the front of the nucleus is compensated by an equal and opposite
orbital angular momentum from a corresponding point on the back of the
nucleus.  This is the resolution to the apparent paradox
\eqref{TMD16}: when we neglect all Wilson line contributions (both
$S_{{\ul x} {\ul y}}[+\infty, b^-]$ and $f_{1T}^{\perp N}$), the net
asymmetry in the quark distribution is zero since $\int d b^-
W_{unp}^{OAM}(p,b) = 0$.  Hence neglecting all Wilson line
contributions yields zero Sivers function, consistent with
\cite{Collins:1992kk}.

An essential role is played in \eqref{TMD15}, then, by the
rescattering factor $S_{{\ul x} {\ul y}}[+\infty, b^-]$.  In the OAM
channel, even though the rescattering $S_{{\ul x} {\ul y}}[+\infty,
b^-]$ is not the source of a preferred transverse direction, without
it the net contribution to the Sivers function from OAM would vanish
after integration over $b^-$, as can be gleaned from the left panel in
\fig{SIDIS_OAM_vs_Sivers}. The rescattering factor $S_{{\ul x} {\ul
    y}}[+\infty, b^-]$ is essential because it introduces shadowing
that breaks this front-back symmetry by screening quarks ejected from
the front of the nucleus more than those ejected the back.  The Sivers
function relevant for SIDIS is therefore more sensitive to OAM from
the back of the nucleus than from the front, which prevents the
complete cancellation of the OAM contribution as in \eqref{TMD16}.

This analysis is strikingly similar to the arguments that historically
established the existence of the Sivers function.  As Collins argued
in \cite{Collins:1992kk}, $PT$-invariance of any hadronic eigenstate
prohibits a preferred direction that can generate the Sivers function.
This is directly reflected in the vanishing of \eqref{TMD16} without
the effects of multiple rescattering.  And as Brodsky, Hwang, and
Schmidt demonstrated in \cite{Brodsky:2002cx}, the rescattering
represented by the semi-infinite Wilson lines breaks this symmetry and
permits a preferred direction for the asymmetry.  Unlike that
calculation, however, here the rescattering does not occur as
color-correlated ``lensing'' due to rescattering on the remnants of
the active quark.  Here the interaction is explicitly
color-decorrelated because the rescattering occurs on many nucleons
whose colors are not correlated.  Despite this difference, the
rescattering effects are still sufficient to break the front-back
symmetry and give rise to a net preferred direction for the asymmetry.



\providecommand{\href}[2]{#2}\begingroup\raggedright\endgroup


\end{document}